\documentclass[
preprint,
 amsmath,amssymb,
 aps, physrev,
]{revtex4-2}

\usepackage{mathtools}
\usepackage{ulem}

\usepackage{tikz}
\usepackage{amsbsy}
\usepackage{bbold}

\usepackage{cancel}
\usepackage[mathscr]{euscript}
\usepackage[usenames, dvipsnames]{xcolor}
\usepackage[hypertexnames=false, plainpages=false, bookmarks, linktocpage=true, bookmarksnumbered, colorlinks, linkcolor=NavyBlue, citecolor=black, filecolor=black, urlcolor=NavyBlue]{hyperref}

\usepackage{graphicx}
\usepackage{dcolumn}
\usepackage{bm}
\usepackage{MnSymbol}

\newcommand{\convolcomma}[0]{\overset{\scriptscriptstyle\otimes}{,}}
\newcommand{\commutator}[2]{ \left[#1, #2\right] }
\newcommand{\convolcommutator}[2]{ \left[#1 \convolcomma #2\right] }
\newcommand{\anticommutator}[2]{ \left\{#1, #2\right\} }
\newcommand{\convolanticommutator}[2]{ \left\{#1 \convolcomma #2\right\} }
\newcommand{\pp}[0]{\mathbf{p}}
\newcommand{\qq}[0]{\mathbf{q}}
\newcommand{\RR}[0]{\mathbf{R}}
\newcommand{\textd}[0]{\mathrm{d}}
\newcommand{\textK}[0]{\mathrm{K}}
\newcommand{\textR}[0]{\mathrm{R}}
\newcommand{\textA}[0]{\mathrm{A}}
\newcommand{\real}[0]{\mathrm{Re}}
\newcommand{\imag}[0]{\mathrm{Im}}
\newcommand{\invGnot}[0]{G^{-1}_{0}}
\newcommand{\GK}[0]{G^{\textK}}
\newcommand{\SigmaK}[0]{\Sigma^{\textK}}
\newcommand{\texteq}[0]{\text{eq}}
\newcommand{\eph}[0]{\text{e-ph}}
\newcommand{\Deltaell}[0]{\text{ }\mathclap{\bar{\Delta}}{\ell}}
\newcommand{\Piell}[0]{\text{ }\mathclap{\bar{\Pi}}{\ell}}

\makeatletter
\newcommand*{\mint}[1]{%
  \mint@l{#1}{}%
}
\newcommand*{\mint@l}[2]{%
  \@ifnextchar\limits{%
    \mint@l{#1}%
  }{%
    \@ifnextchar\nolimits{%
      \mint@l{#1}%
    }{%
      \@ifnextchar\displaylimits{%
        \mint@l{#1}%
      }{%
        \mint@s{#2}{#1}%
      }%
    }%
  }%
}
\newcommand*{\mint@s}[2]{%
  \@ifnextchar_{%
    \mint@sub{#1}{#2}%
  }{%
    \@ifnextchar^{%
      \mint@sup{#1}{#2}%
    }{%
      \mint@{#1}{#2}{}{}%
    }%
  }%
}
\def\mint@sub#1#2_#3{%
  \@ifnextchar^{%
    \mint@sub@sup{#1}{#2}{#3}%
  }{%
    \mint@{#1}{#2}{#3}{}%
  }%
}
\def\mint@sup#1#2^#3{%
  \@ifnextchar_{%
    \mint@sup@sub{#1}{#2}{#3}%
  }{%
    \mint@{#1}{#2}{}{#3}%
  }%
}
\def\mint@sub@sup#1#2#3^#4{%
  \mint@{#1}{#2}{#3}{#4}%
}
\def\mint@sup@sub#1#2#3_#4{%
  \mint@{#1}{#2}{#4}{#3}%
}
\newcommand*{\mint@}[4]{%
  \mathop{}%
  \mkern-\thinmuskip
  \mathchoice{%
    \mint@@{#1}{#2}{#3}{#4}%
        \displaystyle\textstyle\scriptstyle
  }{%
    \mint@@{#1}{#2}{#3}{#4}%
        \textstyle\scriptstyle\scriptstyle
  }{%
    \mint@@{#1}{#2}{#3}{#4}%
        \scriptstyle\scriptscriptstyle\scriptscriptstyle
  }{%
    \mint@@{#1}{#2}{#3}{#4}%
        \scriptscriptstyle\scriptscriptstyle\scriptscriptstyle
  }%
  \mkern-\thinmuskip
  \int#1%
  \ifx\\#3\\\else_{#3}\fi
  \ifx\\#4\\\else^{#4}\fi  
}
\newcommand*{\mint@@}[7]{%
  \begingroup
    \sbox0{$#5\int\m@th$}%
    \sbox2{$#5\int_{}\m@th$}%
    \dimen2=\wd0 %
    \let\mint@limits=#1\relax
    \ifx\mint@limits\relax
      \sbox4{$#5\int_{\kern1sp}^{\kern1sp}\m@th$}%
      \ifdim\wd4>\wd2 %
        \let\mint@limits=\nolimits
      \else
        \let\mint@limits=\limits
      \fi
    \fi
    \ifx\mint@limits\displaylimits
      \ifx#5\displaystyle
        \let\mint@limits=\limits
      \fi
    \fi
    \ifx\mint@limits\limits
      \sbox0{$#7#3\m@th$}%
      \sbox2{$#7#4\m@th$}%
      \ifdim\wd0>\dimen2 %
        \dimen2=\wd0 %
      \fi
      \ifdim\wd2>\dimen2 %
        \dimen2=\wd2 %
      \fi
    \fi
    \rlap{%
      $#5%
        \vcenter{%
          \hbox to\dimen2{%
            \hss
            $#6{#2}\m@th$%
            \hss
          }%
        }%
      $%
    }%
  \endgroup
}
\begin{document}

\title{\textbf{Interband and kinetic corrections to the electronic Boltzmann transport equation} 
}

\author{Elena Trukhan}
\email{Contact author: elena.trukhan@physik.hu-berlin.de}
\author{Nakib H. Protik}%
\email{Contact author: nakib.protik@physik.hu-berlin.de}
\affiliation{%
 Institut f\"{u}r Physik and CSMB, Humboldt-Universit\"{a}t zu Berlin
}

\date{\today}

\begin{abstract}
Interband effects such as coherence/tunneling have recently been shown to give an important contribution to the charge and heat transport properties under certain conditions. These can be captured by adding corrective terms to the semiclassical Boltzmann transport equation. In recent derivations of this type of transport equations that are based on the density matrix formalism, there remain, however, certain omissions. These derivations also rely on a particular type of relaxation time approximation and a band-diagonal form of the interaction self-energies. In this work we derive the interband terms of the electronic Boltzmann transport equation starting from the Keldysh formulation of the quantum kinetic equation and considering the band non-diagonality of the electron-impurity and electron-phonon self-energies. We introduce a minimally modified Kadanoff-Baym \textit{Ansatz}, and find a quantum-corrected, matrix Boltzmann transport equation that is well beyond the current state of the art theory. We show that the occupations and coherences are interdependent, and that the kinetic corrections due to the included interactions cannot, in general, be ignored. This work clarifies the various approximations that must be introduced in an \textit{ab initio} derivation of Boltzmann-like equations and finds a new matrix Boltzmann transport equation that is suitable for parameters-free numerical implementations.
\end{abstract}

\maketitle

\section{Introduction}
The Boltzmann transport equation (BTE), put forth by Ludwig Boltzmann in 1872 \cite{boltzmann1872weitere}, continues to be one of the main theoretical tools for the computation of the charge, heat, and thermoelectric transport properties in condensed matter systems. Over the last decade, the emergence of parameters-free computational methods has enabled solving a semiclassical adaptation of this equation for electrons and phonons in various types of materials, both 3D and 2D \cite{ponce2016epw, PhysRevB.97.121201, ponce2020first, ponce2021first, lee2023electron, bernardi2016first, zhou2021perturbo, PhysRevX.14.021023, nmgj-yq1g, peng2025efficient, li2014shengbte, carrete2017almabte, togo2023implementation, protik2022elphbolt, cepellotti2022phoebe, han2022fourphonon, xiao2025phonon, esfarjani2025fundamentals, mcgaughey2025phonon, claes2025phonon, park2025advances}. The program has seen a decent amount of success. While for certain materials, this methodology has produced transport coefficient pre- and postdictions in good agreement with measurements, for others, it has failed to do so. Broadly speaking, there has been intense research activities in three directions, aimed at improving the accuracy of transport computations. One focuses on the electronic structure aspect, with the goal of improving the band structure calculations. Another focuses on improving the calculation of the interaction self-energies, both the quality of the interaction vertices and the order of the perturbation theory. And the third is focused on improving the Boltzmann transport theory itself. This work falls in the last category. We start here by discussing the salient features of the BTE, highlighting its shortcomings, and motivating the need for a generalization.

In the standard approach, the BTE is the equation of motion of the non-equilibrium occupation function. It is assumed that the interactions in the system are weak, which means that a perfectly sharp relation between the eigenenergies and wavevectors may be assumed. This is called the \textit{quasiparticle} or \textit{on-shell} approximation. It is also assumed that the interactions are local in space and time, the implication being that between successive scattering events, the particles move only under the influence of a non-zero (non-equilibrium condition) external field. This approximation maybe relaxed to capture what are known as \textit{kinetic corrections}, that are the effects of interactions on the action of the external field on the particles. While the external field drives the system out of equilibrium, the effect of the various types of collisions in the system is to relax it back toward equilibrium. The latter is captured by a \textit{collision integral} on the right hand side of the BTE, which gives the balance of the repopulation and depopulation of the single particle eigenstates due to interactions. The construction of the collision integral requires the knowledge of the self-energies, and one cannot set them to zero under the weak interactions assumption that we made earlier. Evidently, doing so will remove any mechanism of attaining a non-zero conductivity -- an experimentally observed fact for normal phase transport. Often times, the collision integral is simplified using phenomenological constructions that go by the umbrella term, relaxation time approximation (RTA). There are several variants of it. In one variant, the repopulation term is droppped. This goes by the name of self-energy relaxation time approximation (SERTA) \cite{lee2023electron}. (To be clear, the software package described in Ref. \cite{lee2023electron} provides both the SERTA and full BTE solvers.) And lastly, the occupation function is taken to be a band-diagonal quantity in the eigenbasis of the non-interacting sector of the Hamiltonian.

All of the above, however, are approximations borrowed and adapted from Boltzmann's original treatment of the non-equilibrium phenomena in the ideal, \textit{classical} gas. But in order to probe certain essential \textit{quantum} effects in transport physics, we necessarily need to go beyond the standard Boltzmann paradigm. We now discuss a few such approaches. For example, the quasiparticle approximation has been relaxed by Castellano and coworkers in the phonon transport problem \cite{castellano2025}. They report that for the 3-phonon limited transport, their theory predicts a significantly different thermal conductivity compared to the corresponding quasiparticle theory of the same class. On a different front, kinetic corrections to the phonon BTE were derived by Horie and Krumhansl \cite{PhysRev.136.A1397} including 3-phonon interactions. In their final expressions, the kinetic corrections show up as a velocity renormalization due to the anharmonicity. Interestingly, they conclude that these effects arise not just from the interactions between different phonon wavepackets (the wavepacket being a particle-like object), but also from the interactions of the waves that constitute a wavepacket. The latter phenomenon has been interpreted as a type of coherence effect by Zhang et al. in Ref. \cite{zhang2022heat}. Meier \cite{meier1969green}, however, commented that the presence of the kinetic corrections in Horie and Krumhansl's work is inconsistent with their invocation of the quasiparticle approximation. These issues of generalizing the BTE to capture crucial quantum physics seemingly apply also to the electronic transport problem.

Now, another notable omission in the standard BTE that has been brought to light in recent times is the interband coherence effect. This has been discussed extensively for the case of phonons by Simoncelli and workers in Refs.  \cite{simoncelli2019unified} and 
\cite{PhysRevX.12.041011}, and for the electronic case by Cepellotti and Kozinsky in Ref. \cite{cepellotti2021interband}. While in the standard BTE, the occupation function is diagonal in the band space, the authors in the above mentioned pioneering works demonstrated that there are off-diagonal terms, interpreted as interband coherences or quantum tunneling effects, that may provide an important contribution to the heat, charge, and thermoelectric transport properties. (We note here that the role of a different type of phonon coherence on electron and phonon transport has been discussed by Stefanucci and Perfetto in Ref. \cite{10.21468/SciPostPhys.16.3.073}. The interband coherence we are discussing here is, to the best of our knowledge, an unrelated phenomenon.) Since these coherence terms are purely quantum in nature, the transport equations derived in these works essentially have quantum corrections added to the otherwise semiclassical BTE. A common theme in these works is that the transport equations are derived starting with the equation of motion of the one-body density matrix. Two other shared features are that (1) \textit{there does not exist any kinetic correction} and (2) \textit{the interband correction to the collision integral is treated in an RTA, following Ref. \cite{PhysRevB.96.115420}}. We will demonstrate later that a consequence of using their RTA in this context is that the occupations and coherences become decoupled. Now, in order go beyond this state of the art theory, we first note that apart from the density matrix formalism, there are other methods for deriving transport theories that provide a more direct insight into the type of approximations that must be made when going from a general quantum kinetic equation to a simpler Boltzmann-like equation. Understanding these approximations allows building generalizations over the standard BTE in a systematic manner.

In this work, we carry out an \textit{ab initio} derivation of a quantum-corrected, linearized, steady state electronic BTE for the direct current (DC) field case, starting from the general quantum kinetic equation in the Keldysh formalism. We consider both the electron-phonon (e-ph) and  electron-impurity (e-imp) interactions and derive interband corrections to both the kinetic and the collision sides. We corroborate the results of the kinetic side of Ref. \cite{cepellotti2021interband} but find additional quantum terms. We find a completely different expression for the interband part of the collision side. Since our transport equation is beyond any type of RTA, it includes both the in- and out-scattering terms. We employ a minimally modified Kadanoff-Baym \textit{Ansatz}. The transport equation we derive is non-diagonal in the band space and, henceforth, we will call it the matrix BTE. We demonstrate the correspondence of our \textit{matrix} BTE to the \textit{standard} BTE in the appropriate limits. 

The main results of this work are as follows.
\begin{enumerate}
    \item Contrary to the prevailing wisdom, there exists a non-vanishing kinetic correction that contains, furthermore, both occupation and coherence dependencies.
    \item Our formalism retains the interband terms in the self-energies.
    \item Our collision integral includes both the impurity and phonon interactions, and captures the coherence effects beyond the current state of the art RTA.
    \item Population and coherence effects are, in general, interdependent.
\end{enumerate}

We present the material in the following manner:

In Sec. \ref{sec:QKE} we present the mathematical notation, the various Green's functions, and the e-ph and e-imp self-energies. We also state the Keldysh quantum kinetic equation in its full form and introduce the Wigner function. Lastly, we explain the two main approximations that we make in this work, namely quasiparticle and weak-field, and motivate our choice of the \textit{Ansatz} solution.

Then, in Sec. \ref{sec:coll}, we give the linearized expressions for the collision integrals for the e-ph and e-imp cases.

After that, in Sec. \ref{sec:kinetic}, we give the linearized expressions for the kinetic side, including the pure field-coupling term and the kinetic corrections due to the interactions.

Then, in Sec. \ref{sec:matrixBTE}, we accumulate the various terms of the working equation, the linearized matrix BTE, giving them in their explicit matrix elemental form.

Lastly, in Sec. \ref{sec:conductivity}, we derive the expressions for the charge current and the conductivity.

We give various proofs and further details of the calculations in the appendices.

\section{Quantum kinetic equation} \label{sec:QKE}
In this section we present the Keldysh quantum kinetic equation (QKE) in the form in which subsequent explicit calculations will be performed. We set $\hbar$ to unity everywhere.

\subsection{Notation}
Before we present the equation, we describe the mathematical notation used here. These have been given in Ref. \cite{rammer2007quantum}, but we reproduce them for completeness.

The center of mass space and time coordinates are defined as
\begin{align}
    (\RR, T) &\equiv \dfrac{1}{2}(\mathbf{r}_{1} + \mathbf{r}_{2}, t_{1} + t_{2}),
\end{align}
where $(\mathbf{r}_{i}, t_{i})$ is the space-time coordinate of the particle at $i$.

The corresponding relative space and time coordinates are
\begin{align}
    (\mathbf{r}, t) &\equiv (\mathbf{r}_{1} - \mathbf{r}_{2}, t_{1} - t_{2}).
\end{align}

For the latter coordinates, we define the reciprocal space duals:
\begin{equation}
    (\mathbf{r}, t) \Leftrightarrow (\mathbf{p}, \omega),
\end{equation}
where $\mathbf{p}$ $(\omega)$ is the momentum (frequency) conjugate to the relative space (time) coordinate.

We use the following 4-vector notation for the following quantities:
\begin{align}
    X &\equiv (T, \RR) \nonumber \\
    x &\equiv (t, \mathbf{r}) \nonumber \\
    p &\equiv (\omega, \mathbf{p}).
\end{align}

Furthermore, we define the corresponding 4-derivatives:
\begin{align}
    \partial^{\Box}_{X} &\equiv (-\partial^{\Box}_{T}, \partial^{\Box}_{\RR}) \nonumber \\
    \partial^{\Box}_{p} &\equiv (-\partial^{\Box}_{E}, \partial^{\Box}_{\mathbf{p}}),
\end{align}
where ${\Box}$ is a placeholder for the operator on which the partial derivative must act, commuting through any other operator, should that be necessary.

Contraction of 4-vectors are defined as follows:
\begin{equation}
    ab \equiv \sum_{\mu\nu = 1}^{4}g_{\mu\nu}a_{\mu}b_{\nu},
\end{equation}
where $a$ and $b$ are 4-vectors and the metric is denoted with $g = \text{diag}\begin{pmatrix} -1 & 1 & 1 & 1 \end{pmatrix}$.

\subsection{Wigner representation, Moyal product, and gradient approximation}
Quantities like $A(X, p)$ are said to be in the mixed or Wigner representation. The product of two objects in this representation is given by \cite{10.1143/PTP.116.61}
\begin{equation}
    (A \otimes B)(X, p) = \exp\left[\dfrac{i}{2}(\partial^{A}_{X}\partial^{B}_{p} - \partial^{A}_{p}\partial^{B}_{X})\right]A(X, p)B(X, p).
\end{equation}
The above is known as the Moyal product.

We will use the so-called gradient approximation, where the Moyal product is truncated up to the first partial derivatives, yielding
\begin{equation}
    (A \otimes B)(X, p) \approx AB + \dfrac{i}{2}\left(\partial_{X}A\partial_{p}B - \partial_{p}A\partial_{X}B\right).
\end{equation}

We will encounter the Moyal commutator and anticommutator later in the work. We precompute these here in the gradient approximation:
\begin{align}
    \convolcommutator{A}{B} &\approx \commutator{A}{B} + \dfrac{i}{2}\commutator{A}{B}_{p} - \dfrac{i}{2}\commutator{B}{A}_{p} \nonumber \\
    \convolanticommutator{A}{B} &\approx \anticommutator{A}{B} + \dfrac{i}{2}\commutator{A}{B}_{p} + \dfrac{i}{2}\commutator{B}{A}_{p},
\end{align}
where we have introduced the generalized Poisson bracket
\begin{equation}
    \commutator{A}{B}_{p} \equiv \partial_{X}A\partial_{p}B - \partial_{p}A\partial_{X}B.
\end{equation}

\subsection{Green's function and other related quantities}
In this formalism, we will encounter Green's functions $G$, self-energies $\Sigma$, spectral functions $A$, and broadening (or linewidth) $\Gamma$. They come in interacting, non-interacting, equilibrium, and non-equilibrium flavors. For the non-interacting case, ``$0$" will be used. And for the equilibrium case, we will use ``$\texteq$". (These ornaments may appear as super- or subscripts.) Some of these quantities come in retarded, advanced, greater, lesser, and Keldysh flavors. These will be denoted with supercripts $\mathrm{R}$, $\mathrm{A}$, $>$, $<$, and $\mathrm{K}$, respectively. The following definitions and relations will be used throughout.
\begin{align}
    A &\equiv i(G^{\mathrm{R}} - G^{\mathrm{A}}) = -2\mathrm{Im}G^{\mathrm{R}} \nonumber \\
    \Gamma &\equiv i(\Sigma^{\mathrm{R}} - \Sigma^{\mathrm{A}}) = -2\mathrm{Im}\Sigma^{\mathrm{R}} \nonumber \\
    \real G &= \dfrac{1}{2}(G^{\mathrm{R}} + G^{\mathrm{A}}) \nonumber \\
    \real\Sigma &= \dfrac{1}{2}(\Sigma^{\mathrm{R}} + \Sigma^{\mathrm{A}}) \nonumber \\
    G^{\mathrm{K}} &\equiv G^{>} + G^{<} \nonumber \\
    \Sigma^{\mathrm{K}} &\equiv \Sigma^{>} + \Sigma^{<}.
\end{align}

All quantities above are defined on the $2 \times 2$ Keldysh contour \cite{Keldysh:1964ud}. Specifically, they are written in the Larkin-Ovchinnikov (upper triangular) form \cite{larkin1975nonlinear}. In this representation, the top-left component is retarded, bottom-right is advanced, and top-right is Keldysh flavored.

\subsection{Electron-impurity and electron-phonon self-energies}\label{ssec:selfens}
We treat the electron-impurity (e-imp) scattering within the first Born approximation:
\begin{equation}\label{eq:sigmaimp}
    \Sigma_{\pp}^{\text{e-imp}}(\omega, \RR) = \sum_{\pp'} V_{\pp \pp'}G_{\pp'}(\omega, \RR)V_{\pp'\pp}^\text{T},
\end{equation}
where $V_{\pp \pp'}$ is the interaction vertex which has the property $V_{\pp' \pp} = V^{*}_{\pp \pp'}$ \cite{ashcroft1976solid, bruus_manybody_2004}. In the eigenbasis of the non-interacting Hamiltonian $H_{\pp}^{0, \texteq}$, $\Sigma_{\pp}^{\text{e-imp}}$ is a generally non-diagonal square matrix. Above, $G$ should be understood as a generic Green's function, the flavor of which (R, A, or K) matches that of $\Sigma$ \cite{RevModPhys.58.323}.

The electron-phonon (e-ph) self-energy is quite a bit more complex. Generalizing Eq. 7.111 of Ref. \cite{rammer2007quantum} (which takes the interaction to be a constant) with the momenta and electron and phonon band resolved Fan-Migdal expression, Eq. 157 of Ref. \cite{giustino2017electron}, we get
\begin{equation}
    \Sigma^{\eph}_{\pp ij}(\omega, \RR) = i\sum_{i'j'k'k}\gamma_{ii'k}\sum_{s}\sum_{\pp'}\int\dfrac{\textd{\omega'}}{2\pi}\dfrac{g^{\dagger}_{s\pp\pp'}}{\sqrt{\omega_{s\qq}/2}}G_{\pp'i'j'}(\omega', \RR)\dfrac{g_{s\pp\pp'}}{\sqrt{\omega_{s\qq}/2}}D_{s\qq kk'}(\omega - \omega', \RR)\tilde{\gamma}_{j'jk'}.
\end{equation}

Above, the indices $i, j, k$ are $\in [1, 2]$. The e-ph interaction is given by $g$, which is a square matrix in the electronic band space and is tagged by the phonon band $s$. This matrix element is scaled by the factor $\sqrt{\omega_{s\qq}/2}$, which is done to reflect the convention used to define the phonon Green's function here (this will defined shortly). The electron and phonon contour Green's functions are denoted with $G$ and $D$, respectively. The wave vector transferred due to the interaction is $\qq \equiv \pp' - \pp$. The rank-3 tensors $\gamma$ and $\tilde{\gamma}$ are called the phonon absorption and emission ``vertices" \cite{RevModPhys.58.323}, respectively, and are defined in the following manner:
\begin{align}
    \gamma_{ij1} &= \dfrac{1}{\sqrt{2}}\mathbb{1}_{ij}, \quad 
    \gamma_{ij2} = \dfrac{1}{\sqrt{2}}\tau^{1}_{ij} \nonumber \\
    \tilde{\gamma}_{ij1} &= \dfrac{1}{\sqrt{2}}\tau^{1}_{ij}, \quad \tilde{\gamma}_{ij2} = \dfrac{1}{\sqrt{2}}\mathbb{1}_{ij}
\end{align}

Above $\mathbb{1}$ is the identity matrix and $\tau^{1} = \text{off-diag}(1 \quad 1)$ is the first Pauli matrix.

We carry out the contour summations and obtain the expressions for the various components of the phonon self-energy. We first define the shorthand $\Lambda_{ij} \equiv g^{\dagger}G_{ij}g$, suppressing other dependencies for brevity. Then, carrying out the contractions by hand, we get
\begin{equation}
    \Sigma^{\eph}_{\pp}(\omega, \RR) = \dfrac{i}{2}\sum_{s}\sum_{\pp'} \dfrac{1}{\omega_{s\qq}/2} \begin{bmatrix}
        \Lambda^{\text{K}}\star D^{\text{R}} + \Lambda^{\text{R}}\star D^{\text{K}} & \Lambda^{\text{R}}\star D^{\text{R}} + \Lambda^{\text{K}}\star D^{\text{K}} + \Lambda^{\text{A}}\star D^{\text{A}} \\
        \Lambda^{\text{A}}\star D^{\text{R}} + \Lambda^{\text{R}}\star D^{\text{A}} & \Lambda^{\text{A}}\star D^{\text{K}} + \Lambda^{\text{K}}\star D^{\text{A}},
    \end{bmatrix}
\end{equation}
where the $\star$ product denotes the convolution
\begin{equation}
    A(\omega) \star B(\omega) \equiv \int\dfrac{\textd\omega'}{2\pi}A(\omega')B(\omega - \omega').
\end{equation}

The bottom left element of $\Sigma^{\eph}$ vanishes because products of retarded and advanced Green's functions are trivially zero. In fact, using this identity and the fact that $D$ commutes with both $g$ and $G$ (since the phononic and electronic band spaces are independent), we can bring the top-right (Keldysh) component into a form that is more convenient for computation. Doing so, we get
\begin{equation}\label{eq:sigmaeph}
    \Sigma^{\eph}_{\pp}(\omega, \RR) = \dfrac{i}{2}\sum_{s}\sum_{\pp'} \dfrac{1}{\omega_{s\qq}/2} \begin{bmatrix}
        \Lambda^{\text{K}}\star D^{\text{R}} + \Lambda^{\text{R}}\star D^{\text{K}} & (\Lambda^{\text{R}} - \Lambda^{\text{A}})\star (D^{\text{R}} - D^{\text{A}}) + \Lambda^{\text{K}}\star D^{\text{K}} \\
        0 & \Lambda^{\text{A}}\star D^{\text{K}} + \Lambda^{\text{K}}\star D^{\text{A}}
    \end{bmatrix}
\end{equation}

An important point to note is that the impurity and phonon interaction matrix elements are assumed to be static (no $\omega$ dependence). This is a standard approximation, as discussed in Ref. \cite{giustino2017electron} in the context of phonon scattering. The type -- R, A, or K -- of the self-energy enters only via the type of the Green's function. Also, the impurity interaction matrix elements are assumed to be independent of the phonon ones, and \textit{vice versa}. In other words, we work within an approximation where Matthiesen's rule \cite{ashcroft1976solid} holds. Next, the matrix elements are assumed to be independent of the external driving field. And lastly, we will use the dragless approximation. That is to say, we will assume that the phonon system remains in equilibrium even when the electron system is not, despite the fact that we have explicit electron-phonon coupling. This means that the phonon Keldysh Green's function will always retain its equilibrium form. We will further take the phonon Green's function to be the non-interacting one. Under these approximations, the components of the phonon contour Green's function are
\begin{align}
    &D_{11, s\qq}^{0, \texteq} \equiv D^{0, \text{R}, \texteq}_{s\qq}(\omega) = -\dfrac{\omega^{2}_{s\qq}}{\omega^{2}_{s\qq} - (\omega + i0^{+})^{2}} \nonumber \\
    &D_{22, s\qq}^{0, \texteq} \equiv D^{0, \text{A}, \texteq}_{s\qq}(\omega) = -\dfrac{\omega^{2}_{s\qq}}{\omega^{2}_{s\qq} - (\omega - i0^{+})^{2}} \nonumber \\
    &D_{12, s\qq}^{0, \texteq} \equiv D^{0, \text{K}, \texteq}_{s\qq}(\omega) = -iB_{s\qq}^{0, \texteq}(\omega)l^{\texteq}_{s\qq},
\end{align}
where the last equation is the phonon fluctuation-dissipation theorem and $l^{\texteq}_{s\qq}$ is the equilibrium phonon distribution function which takes the form of
\begin{equation}\label{eq:phononl}
    l^{\texteq}(\omega) = \coth\left(\dfrac{\beta\omega}{2}\right).
\end{equation}
$B^{0, \texteq}$ is the equilibrium, non-interacting phonon spectral function given by
\begin{equation}
    B^{0, \texteq}_{s\qq}(\omega) \equiv i(D^{0, \text{R}, \texteq}_{s\qq}(\omega) - D^{0, \text{A}, \texteq}_{s\qq})(\omega) = \pi\omega_{s\qq}\text{sgn}(\omega)\left[\delta(\omega - \omega_{s\qq}) + \delta(\omega + \omega_{s\qq})\right]
\end{equation}

Crucially, though, we go beyond the standard set of assumptions by considering the full non-diagonality of the self-energies.

\subsection{Keldysh quantum kinetic equation}
In terms of the quantities presented above, the general electronic quantum kinetic equation in the Keldysh formalism is \cite{RevModPhys.58.323} 
\begin{align} \label{eq:qke}
    \overbrace{\convolcommutator{G^{-1}_{0}}{G^{\textK}}}^{\text{pure field coup.}} \overbrace{- \convolcommutator{\real\Sigma}{G^{\textK}}}^{\text{kinetic corr. 1}} \overbrace{- \convolcommutator{\Sigma^{\textK}}{\real G}}^{\text{kinetic corr. 2}} = 
    \overbrace{\dfrac{i}{2}\convolanticommutator{\Sigma^{\textK}}{A}}^{\text{in-scatter.}} \overbrace{- \dfrac{i}{2}\convolanticommutator{\Gamma}{G^{\textK}}}^{\text{out-scatter.}},
\end{align}
where the inverse, bare Green's function is given by
\begin{equation}
    G_{0}^{-1}(\omega, \pp, \RR) = \omega - H^{0}_{\pp} - U(\RR), 
\end{equation}
with $U(\RR)$ being the electric scalar field which drives the electronic system out of equilibrium. Note that in the eigenbasis of $H^{0}_{\pp}$, the above is not a diagonal matrix due to the presence of the driving term.

Eq. \ref{eq:qke} is the starting point for all of our calculations. The first term on the left hand side (LHS) of this equation describes the evolution of the system purely due to a direct coupling to the external field. The remaining two terms on the LHS give the kinetic corrections due to interactions. We identify the right hand side (RHS) terms as the in-scattering (repopulation) and out-scattering (depopulation) parts of the collision integral. In sections \ref{sec:coll} and \ref{sec:kinetic}, we will work on each term separately. For all terms, we will apply the gradient approximation in the expansion of the Moyal product. Then, following the evaluation of the Poisson brackets, we will integrate out the frequency dependence. And lastly, we will express everything in the eigenbasis of $H_{\pp}^{0}$. The last two steps commute, as shown in Appendix Sec. \ref{appsec:basischange}. This workflow will yield a Boltzmann-like transport equation in terms of the matrix valued Wigner function $h_{\pp}$ defined as (also see Appendix Sec. \ref{appsec:wigner})
\begin{equation} \label{eq:wignerfn}
    h_{\pp} \equiv -\int\dfrac{\textd\omega}{2\pi i}\GK_{\pp}(\omega).
\end{equation}

The workflow above generalizes the one given in Ref. \cite{rammer2007quantum} for non-diagonal self-energies and Wigner function. Recall that the diagonal part of the Wigner function can be interpreted as the population (or occupation) of the state, whereas the off-diagonal part gives the interband coherence effects \cite{PhysRevX.12.041011} at a given wave vector.

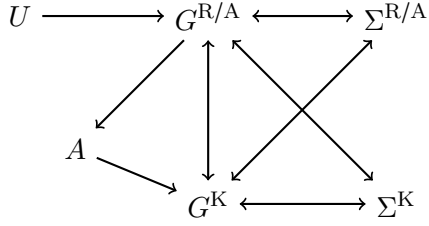
\begin{figure}[ht!]
\centering
\begin{tikzpicture}[node distance={25mm}, thick, main/.style = {}]
\node[main] (1) {$U$};
\node[main] (2) [right of=1] {$G^{\text{R}/\text{A}}$}; 
\node[main] (3) [right of=2] {$\Sigma^{\text{R}/\text{A}}$}; 
\node[main] (4) [below left of=2] {$A$}; 
\node[main] (5) [below of=3] {$\Sigma^{\text{K}}$}; 
\node[main] (6) [below of=2] {$G^{\text{K}}$}; 
\draw[->] (1) to (2); 
\draw[<->] (2) to (3); 
\draw[<->] (2) to (6);
\draw[->] (2) to (4);
\draw[->] (4) to (6);
\draw[<->] (6) to (3);
\draw[<->] (6) to (5);
\draw[<->] (5) to (2);
\end{tikzpicture} 
\caption{Dependencies between the Green's functions, self-energies, spectral function, and external field in the general QKE. The arrow shows injection of dependency.}
\label{fig:dependency-full}
\end{figure}

\subsection{Quasiparticle and weak-field approximations and \textit{Ansatz} solution}
We restrict ourselves to the weak-field transport case. Furthermore, we assume that the energy renormalization effects of interactions are negligible. By these we specifically mean that we ignore both the external field and the interaction self-energy in the spectral function. This leads to a tremendous simplification over the recursive functional dependencies between the various quantities. That is,
\begin{equation}\label{eq:quasiparticle-weakfield}
    A[G[\Sigma[G]]] \approx A^{0, \texteq}[G^{0, \texteq}] = 2\pi\delta(\omega - H^{0}_{\pp}) \Rightarrow 2\pi\delta(\omega - \epsilon_{\pp}).
\end{equation}
Note that within these approximations, the spectral function in the eigenbasis of $H_{\pp}^{0}$ becomes a diagonal matrix. We will see shortly that this allows simplifying the various terms into a compact form. The basis change is shown with the $\Rightarrow$. The quantity $\epsilon_{\pp}$ is a diagonal matrix of the eigenenergies at wavevector $\pp$, measured with respect to the chemical potential.

The application of these approximations to the QKE is somewhat subtle due to the fact that the various quantities are interconnected in a complex manner. The functional dependencies for the most general case are shown in Fig. \ref{fig:dependency-full}.

\begin{figure}[ht!]
\centering
\begin{tikzpicture}[node distance={25mm}, thick, main/.style = {}]
\node[main] (1) {$U$};
\node[main] (2) [right of=1] {$G^{0,\text{R}/\text{A}}$}; 
\node[main] (3) [right of=2] {$\Sigma^{0,\text{R}/\text{A}}$}; 
\node[main] (4) [below left of=2] {$A^{0}$}; 
\node[main] (5) [below of=3] {$\Sigma^{\text{K}}$}; 
\node[main] (6) [below of=2] {$G^{\text{K}}$}; 
\draw[->] (1) to (2); 
\draw[->] (2) to (3); 
\draw[->] (2) to (6);
\draw[->] (2) to (4);
\draw[->] (4) to (6);
\draw[<->] (6) to (3);
\draw[<->] (6) to (5);
\draw[->] (2) to (5);
\end{tikzpicture} 
\caption{Dependencies between the Green's functions, self-energies, spectral function, and external field in the QKE after applying the quasiparticle approximation.}
\label{fig:dependency-qp}
\end{figure}
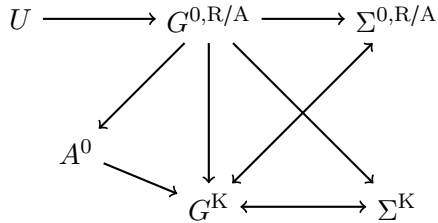

Applying just the quasiparticle approximation, i.e. ignoring the self-energy effects in just the retarded/advanced quantities, the functional dependencies simplify significantly. This is because the self-consistency between $G^{\text{R/A}}$ and $\Sigma^{\text{R/A}}$ is no longer required. This is shown in Fig. \ref{fig:dependency-qp}. The situation still remains complex since the presence of the external field $U$ in the retarded/advanced Green's function generally renders these and also the spectral function non-diagonal in the eigenbasis of $H^{0}$. 

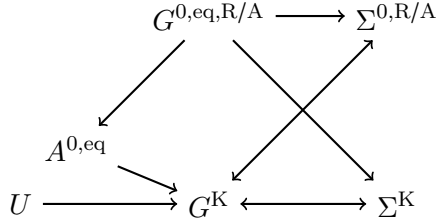
\begin{figure}
\centering\begin{tikzpicture}[node distance={25mm}, thick, main/.style = {}]
\node[main] (2) {$G^{0, \texteq, \text{R}/\text{A}}$}; 
\node[main] (3) [right of=2] {$\Sigma^{0,\text{R}/\text{A}}$}; 
\node[main] (4) [below left of=2] {$A^{0, \texteq}$}; 
\node[main] (5) [below of=3] {$\Sigma^{\text{K}}$}; 
\node[main] (6) [below of=2] {$G^{\text{K}}$}; 
\node[main] (1) [left of=6] {$U$};
\draw[->] (2) to (3); 
\draw[->] (2) to (4);
\draw[->] (4) to (6);
\draw[<->] (6) to (3);
\draw[<->] (6) to (5);
\draw[->] (2) to (5);
\draw[->] (1) to (6);
\end{tikzpicture} 
\caption{Dependencies between the Green's functions, self-energies, spectral function, and external field in the QKE after applying the quasiparticle and weak-field approximations, and the \textit{Ansatz}.}
\label{fig:dependency-qp-weak-field-ansatz}
\end{figure}

Next, we invoke the weak-field approximation which allows these quantities to be diagonal. However, the $U$-dependence of $\GK$ now has to be introduced using a different trick. This is illustrated in Fig. \ref{fig:dependency-qp-weak-field-ansatz}. In this set of approximations, the effect of $U$ is to take $G^{\text{K}, \texteq}$ to $\GK$. And this is captured in two ways. First, $U$ is retained in the Moyal commutator between $G^{-1}_{0}$ and $\GK$ (first term on the LHS of Eq. \ref{eq:qke}). It is important to retain this since the Poisson bracket will furnish a linear order $\partial_{\RR}U$ factor, which defines the electric field $\mathbf{E} \equiv \partial_{\RR}U/e$, where $e$ is the magnitude of the electronic charge. And the second way the effect of the field is captured is via an \textit{Ansatz} solution
\begin{equation}\label{eq:ansatz}
    \GK_{\pp} \doteq -\dfrac{i}{2}\{A_{\pp}, h_{\pp}\} \approx -\dfrac{i}{2}\{A_{\pp}^{0, \texteq}, h_{\pp}\},
\end{equation}
where $h_{\pp}$ is the non-equilibrium Wigner function. The Wigner function, by construction, is a Hermitian matrix. (Note that, by definition, $\GK$ is antihermitian.) We will further retain the self-energies in the rest of the LHS terms as these give $\mathscr{O}(\mathbf{E})$ kinetic corrections.

Note that in equilibrium, Eq. \ref{eq:ansatz} above is an exact result, namely, the fluctuation-dissipation theorem \cite{RevModPhys.58.323} where the equilibrium form of $h$ is
\begin{equation}\label{eq:electronh}
    h^{\texteq}(\omega) = \tanh\left(\dfrac{\beta\omega}{2}\right).
\end{equation}
The exactness of Eq. \ref{eq:ansatz} in this case follows from the fact that in equilibrium, all quantities are band diagonal. In the literature \cite{Kadanoff, RevModPhys.58.323}, however, one encounters the Kadanoff-Baym \textit{Ansatz} (KBA): $\GK_{\pp} \doteq -iA_{\pp}h_{\pp}$. This is suitable for single band cases, or in the particular multiband case where both $A$ and $h$ can be safely approximated as diagonal matrices. But for a general multiband system where interband effects are important, the KBA cannot be used since $h_{\pp}$ and $A_{\pp}$ (or even $A_{\pp}^{0, \texteq}$, for that matter) do not commute, and, as a consequence, the antihermiticity of $\GK_{\pp}$ is destroyed. A variant of the KBA is the free generalized KBA, which proposes that $\GK_{mn\pp} \doteq -i2\pi \delta\left(\omega - (\epsilon_{m\pp} + \epsilon_{n\pp})/2\right)h_{mn\pp}$ \cite{unlu2004multi}. We, however, find it difficult to see under what conditions taking an average of the eigenenergies of two bands in the delta function is justified. Our proposed \textit{Ansatz} minimally generalizes the standard KBA for the multiband case, preserving the fact that the Wigner function must be Hermitian.

Returning to the original question, the \textit{Ansatz} solution allows capturing the effect of the scalar field $U$ since $h_{\pp} = h_{\pp}(U) = h_{\pp}(\mathbf{E})$. In this work, we will work within a linear response regime such that $h_{\pp} \doteq h_{\pp}^{\texteq} + \delta h_{\pp}(\mathbf{E})$, the second term measuring the deviation from equilibrium and being linear in the field $\mathbf{E}$.

Following the application of the quasiparticle and weak-field approximations, and using the \textit{Ansatz} solution, the self-consistency loops between $\GK$ and the self-energies are simple in the sense that there aren't multiple coupled equations of motion to solve. The QKE will now furnish a single integrodifferential equation for $h_{\pp}$, which is the only remaining unknown. This final equation of motion of $h_{\pp}$ would be a quantum-corrected electronic BTE capturing both the effects of the non-diagonality of the self-energies and also the interband coherences/tunnelings. This is what we call the matrix BTE.

\section{Collision terms} \label{sec:coll}
In this section, we present the collision side of our transport equation. Details of the derivation are given in Appendix Sec. \ref{appsec:collision_eph}. The in-scattering term takes the form of
\begin{equation}
    \dfrac{i}{2}\convolanticommutator{\Sigma^{\textK}}{A} \approx \dfrac{i}{2}\convolanticommutator{\Sigma^{\textK}}{A^{0, \texteq}} = \dfrac{i}{2}\anticommutator{\SigmaK}{A^{0, \texteq}} - \dfrac{1}{4}\commutator{\SigmaK}{A^{0, \texteq}}_{p} - \dfrac{1}{4}\commutator{A^{0, \texteq}}{\SigmaK}_{p}.
\end{equation}

In the expression above, we may drop the Poisson brackets since $\partial_{\RR}A^{0, \texteq} = 0$ in the quasiparticle + weak-field approximation, and $\partial_{\RR}\SigmaK = 0$ for the DC field case that we are considering. Also, as mentioned earlier, we only consider the steady state case, which means all $\partial_{T}$ terms vanish.

The out-scattering term is
\begin{equation}
    - \dfrac{i}{2}\convolanticommutator{\Gamma}{G^{\textK}} \approx -\dfrac{i}{2}\convolanticommutator{\Gamma^{0, \texteq}}{\GK} = -\dfrac{i}{2}\anticommutator{\Gamma^{0, \texteq}}{\GK} + \dfrac{1}{4}\commutator{\Gamma^{0, \texteq}}{\GK}_{p} + \dfrac{1}{4}\commutator{\GK}{\Gamma^{0, \texteq}}_{p},
\end{equation}
where, again, the Poisson brackets can be ignored.

Next, we work these terms out separately for the electron-phonon and electron-impurity collision cases and present them in a linearized form, i.e. retaining only up to $\mathscr{O}(\delta h)$ terms.

\subsection{Electron-phonon scattering}

\paragraph*{e-ph in-scattering term}
First we define $\Delta^{\text{emi}, s-\qq}_{\pp\pp'}$ as the matrix of $\delta(\epsilon_{n\pp} - \epsilon_{n'\pp'} - \omega_{s-\qq})$ and $\Delta^{\text{abs}, s\qq}_{\pp\pp'}$ as the matrix of $\delta(\epsilon_{n\pp} - \epsilon_{n'\pp'} + \omega_{s\qq})$. Furthermore, we define $\bar{\Delta}^{s}_{\pp\pp'} \equiv \Delta^{\text{emi}, s-\qq}_{\pp\pp'} -\Delta^{\text{abs}, s\qq}_{\pp\pp'}$ and  $\Deltaell^{s}_{\pp\pp'} \equiv \Delta^{\text{emi}, s-\qq}_{\pp\pp'}l^{\texteq}_{s-\qq}(\omega_{s-\qq}) -\Delta^{\text{abs}, s\qq}_{\pp\pp'} l^{\texteq}_{s\qq}(-\omega_{s\qq})$. The superscripts ``emi" and ``abs" stand for phonon emission and absorption, respectively. Then, using the expression for the Keldysh component of the self-energy from Eq. \ref{eq:sigmaeph}, the linearized in-scattering term turns out to be
\begin{align} \label{eq:ephcollin}
     \dfrac{i}{2}\anticommutator{\SigmaK}{A^{0, \texteq}} &\rightarrow \dfrac{i\pi}{4} \sum_{s\pp'}\Bigg\{ \left[ (g_{s\pp\pp'}^\dagger \delta h_{\pp'}) \odot \Deltaell_{\pp\pp's} \right]g_{s\pp\pp'}  + \left[ g_{s\pp\pp'}^\dagger \odot \Deltaell_{\pp\pp's} \right] \delta h_{\pp'}g_{s\pp\pp'} 
     + h. c.\Bigg\},
\end{align}
where the $h. c.$ is to be replaced by the preceding terms in their Hermitian conjugate forms, and the $\rightarrow$ symbolizes taking the $-\int\textd\omega/(2\pi i)$, followed/preceded by changing to the eigenbasis of $H^{0}_{\pp}$.

\paragraph*{e-ph out-scattering term}
Similarly, using the definition of the linewidth, the linearized out-scattering term evaluates to
\begin{align}\label{eq:ephcollout}
    &-\dfrac{i}{2}\anticommutator{\Gamma^{0}}{\GK}  \rightarrow  \nonumber\\
    &\quad-\dfrac{i\pi}{4}\sum_{s\pp'}\Bigg\{ \biggl( \left[ (\delta h_{\pp}g_{s\pp\pp'}^\dagger )\odot (\bar{\Delta}_{\pp\pp's}h^{\texteq}_{\pp'} +  \Deltaell_{\pp\pp's})\right] + \delta h_{\pp} \left[ g_{s\pp\pp'}^\dagger \odot (\bar{\Delta}_{\pp\pp's} h^{\texteq}_{\pp'} + \Deltaell_{\pp\pp's})\right] \nonumber\\
   &\qquad \qquad  + \left[(h^{\texteq}_{\pp}g_{s\pp\pp'}^\dagger ) \odot \bar{\Delta}_{\pp\pp's}\right]\delta h_{\pp'} + \left[ (h^{\texteq}_{\pp}g_{s\pp\pp'}^\dagger \delta h_{\pp'})\odot \bar{\Delta}_{\pp\pp's} \right] \biggl)g_{s\pp\pp'}  + h. c. \Bigg\}
\end{align}

To write these expressions in the linearized form we used the detailed balance condition, Appendix Sec. \ref{appsec:detbalance}.

We note that in order to get a purely band-diagonal contribution from these terms, we need to ignore the off-diagonal terms of both the self-energy \cite{giustino2017electron} and the Wigner function, $h_{\pp}$. That is, the usual diagonal form of the standard semiclassical BTE requires both these approximations, simultaneously.

A comment is in order regarding the use of RTAs in the pioneering works in this field in Refs. \cite{PhysRevX.12.041011} and \cite{cepellotti2021interband}. In these works, the diagonal part of the collision integral is treated in the SERTA in the numerical calculations (i.e. the in-scattering part is dropped), although the full expression is provided. Next, the off-diagonal part in these works is treated in a different type of RTA, as proposed in Eq. 31 of Ref. \cite{PhysRevB.96.115420}. It resembles a generalized Bhatnagar-Gross-Krook form \cite{bhatnagar1954model}, but introduces a non-diagonal self-energy by demanding that the matrix elements of the linewidths are $\Gamma_{nn'} = (\Gamma_{n} + \Gamma_{n'})/2$. We do not see under what conditions such a construction will generally hold. It is also not easy to see how such a relation can be reached from the \textit{ab initio} formalism employed in this work. As such, compared to the current state of the art theories, our theory provides a completely different non-diagonal collision integral, which is derived entirely from \textit{ab initio} concepts, and is beyond any type of RTA.

\subsection{Electron-impurity scattering}
The evaluation of the e-imp collision terms proceeds as in the e-ph case. Using the e-imp self-energy expression Eq. \ref{eq:sigmaimp}, the in- and out-scattering contributions that result are given below; see Appendix Sec. \ref{appsec:collision_eimp} for the detailed derivation.

\paragraph*{e-imp in-scattering term:}
\begin{align}
     \dfrac{i}{2}\anticommutator{\SigmaK_\pp(\omega)}{A_\pp^{0, \texteq}(\omega)} & \rightarrow  \dfrac{i\pi }{2} \sum_{\pp'}\left( \left[V_{\pp\pp'} \odot \Delta_{\pp\pp'}\right]\delta h_{\pp'}V_{\pp\pp'}^\dagger + \left[(V_{\pp\pp'}\delta h_{\pp'}) \odot \Delta_{\pp\pp'}\right]V_{\pp\pp'}^\dagger  + h.c. \right)
\end{align}

\paragraph*{e-imp out-scattering term:}
\begin{align}
     -\dfrac{i}{2}\anticommutator{\Gamma^{0}_\pp(\omega)}{\GK_\pp(\omega)}& \rightarrow  -\dfrac{i\pi }{2} \sum_{\pp'}\left(\left[\Delta_{\pp\pp'} \odot (\delta h_{\pp}V_{\pp\pp'})\right]V_{\pp\pp'}^\dagger + \delta h_{\pp}\left[\Delta_{\pp\pp'} \odot V_{\pp\pp'}\right]V_{\pp\pp'}^\dagger + h.c. \right)
\end{align}

Similar observations, as made earlier about the e-ph collision terms, hold here too. The collision integral we derived here is beyond any type of RTA. We also note that in the current state of the art theories, impurity scattering has not been considered.

\section{Kinetic terms} \label{sec:kinetic}

Here we derive the LHS of the QKE in a linearized form. Each term gives four combinations of the equilibrium part and the non-equilibrium corrections to the various objects of interest. We keep only pairs consisting of one equilibrium and one non-equilibrium correction, because fully equilibrium terms are canceled due to the detailed balance condition, Appendix Sec. \ref{appsec:detbalance}. And fully non-equilibrium pairs give $\mathscr{O}\left(\mathbf{E}^{2}\right)$ corrections. We work out the direct field coupling and the kinetic correction terms below.

\paragraph*{Direct field coupling term}
The direct field coupling term, after applying the gradient approximation, is
\begin{align}  
    \convolcommutator{G^{-1}_{0}}{G^{\textK}} &\approx  \commutator{ G^{-1, \texteq}_{0}}{\delta G^{\textK}} + \dfrac{i}{2}\commutator{G^{-1, \texteq}_{0}}{\delta G^{\textK}}_{p} - \dfrac{i}{2}\commutator{\delta G^{\textK}}{G^{-1, \texteq}_{0}}_{p} \nonumber \\
    & + \commutator{\delta G^{-1}_{0}}{G^{\textK, \texteq}} + \dfrac{i}{2}\commutator{\delta G^{-1}_{0}}{G^{\textK, \texteq}}_{p} - \dfrac{i}{2}\commutator{G^{\textK, \texteq}}{\delta G^{-1}_{0}}_{p} , 
\end{align}
where $G^{-1, \texteq}_{0} = \omega - H_\pp^0$, $\delta G^{-1}_{0} = - U(\RR)$, $G^{\textK, \texteq}_\pp = -ih^\texteq_\pp  A^{0, \texteq}_\pp$, $\delta G^{\textK, \texteq}_\pp = -\dfrac{i}{2}\anticommutator{\delta h_\pp }{A^{0, \texteq}_\pp}$.

In the homogeneous case, the Poisson brackets associated with the second and third terms can be neglected. The contribution from the fifth and sixth terms is proportional solely to $\partial_\pp h_\pp^\texteq$, as the $\omega$-integral of $\partial_\pp A^{0, \texteq}_\pp$ vanishes (see Appendix \ref{appsec:directfield}). As a result, this term evaluates to
\begin{align} \label{eq:direct_field_coupling}
    \convolcommutator{G^{-1}_{0}}{G^{\textK}} \rightarrow 
    \commutator{\delta h_{\pp}}{\varepsilon_{\pp}} + [h_\pp^\texteq, U(\RR)]  - \dfrac{i}{2}\anticommutator{\partial_{\RR}U(\RR)}{\partial_{\pp}h_{\pp}^\texteq}.
\end{align}

\paragraph*{Kinetic correction terms}
As with the collision terms, all Poisson brackets vanish for a homogeneous electric field. Using the approximation that energy renormalization from the external field is negligible, we can omit further terms containing $\delta\real G$. The term with $\delta \real\Sigma$ can be neglected if there is no contribution to the self-energy from the Keldysh Green's function. This is the case for the e-imp scattering but not for the e-ph one \cite{RevModPhys.58.323}, as discussed in Sec. \ref{ssec:selfens}.

As a result, the two kinetic correction terms are given by the following expressions, respectively: 
\begin{align}
\convolcommutator{\real\Sigma}{G^{\textK}}  & \rightarrow   -\dfrac{1}{2\pi i}\int \textd \omega \left(\commutator{\real \Sigma^\texteq (\omega)}{\delta G^{\textK}(\omega)} +  \commutator{\delta\real  \Sigma (\omega)}{G^{\textK, \texteq}(\omega)} \right) 
\end{align}
and
\begin{align}
\convolcommutator{\SigmaK}{\real G}  & \rightarrow   -\dfrac{1}{2\pi i}\int \textd \omega \commutator{\delta \SigmaK (\omega)}{\real G^{\texteq}(\omega)}.
\end{align}

Further simplification requires knowledge of the exact form of the self-energies, $\Sigma = \Sigma[G]$. Below we evaluate the kinetic correction terms for the e-ph and e-imp cases separately.

\subsection{Electron-phonon kinetic corrections}

In this case, $\real \Sigma$ is given by the following expression:
\begin{align}\label{eq:real_sigma_eph}
    \real \Sigma = \dfrac{1}{2}(\Sigma^\textR + \Sigma^\textA) 
        & = \dfrac{i}{2}\sum_{s\pp'}\dfrac{1}{\omega_{s\qq}} (\Lambda^{\text{K}}\star(D^{\text{R}} + D^{\text{A}}) + (\Lambda^{\text{R}}+\Lambda^{\text{A}})\star D^{\text{K}})
\end{align}

As we can see, there is a contribution from $\Lambda^K$, which is proportional to $\delta h_\pp$, so in the case of the e-ph coupling $\delta\real \Sigma$ can not be neglected. Substitution of the $\Sigma$ given by the formula in Eq. \ref{eq:sigmaeph} leads to the following form of the kinetic terms (see derivations in Appendix Sec. \ref{appsec:kineticcor}):
\begin{align}
    \commutator{\real \Sigma^{\texteq}}{\delta\GK} &\rightarrow -\dfrac{1}{4 } \sum_{s\pp'}\Bigl(\left[ (\delta h_\pp g_s^\dagger) \odot(\Piell_{s\pp\pp'} + \bar{\Pi}_{s\pp\pp'}h_{\pp'}^{\texteq}) \right]g_s \nonumber \\
    & + \delta h_\pp\left[  g_s^\dagger \odot(\Piell_{s\pp\pp'} + \bar{\Pi}_{s\pp\pp'}h_{\pp'}^{\texteq}) \right]g_s - h.c.\Bigl), \\
    \commutator{\delta \real \Sigma}{G^{\textK, \texteq}} &\rightarrow -\dfrac{1}{4} \sum_{s\pp'} \Bigl( \left[(h_{\pp}^{\texteq}g_s^\dagger) \odot \bar{\Pi}_{s\pp\pp'} \right]\delta h_{\pp'}g_s + \left[(h_{\pp}^{\texteq}g_s^\dagger\delta h_{\pp'})\odot\bar{\Pi}_{s\pp\pp'}\right]g_s -  h.c.\Bigl),\\
    \commutator{\delta \SigmaK}{\real G^{\texteq}} &\rightarrow \dfrac{1}{4 }\sum_{s\pp'}\left( \left[ (g_s^\dagger \delta h_{\pp'})\odot \Piell_{s\pp\pp'}\right]g_s - h.c.\right),
\end{align}
where $\Piell_{s\pp\pp'} = \Pi_{\pp\pp'}^{\text{emi}, s-\qq}l^{0, \texteq}_{-\qq}(\omega_{s-\qq})- \Pi_{\pp\pp'}^{\text{abs}, s\qq}l^{0, \texteq}_\qq(-\omega_{s\qq})$, $\bar{\Pi}_{s\pp\pp'} = \Pi_{\pp\pp'}^{\text{emi}, s-\qq}- \Pi_{\pp\pp'}^{\text{abs}, s\qq}$. The emission and absorption $\Pi$ values are given by the following expressions:
\begin{align}
[\Pi_{\pp\pp'}^{\text{emi}/\text{abs},s\mp\qq}]_{nn'} \equiv \begin{cases} 0,  \text{ if } \epsilon_{n\pp} = \epsilon_{n'\pp'} \pm\omega_{s\mp\qq}\\ 
   \dfrac{1}{\epsilon_{n\pp} - (\epsilon_{n'\pp'}\pm\omega_{s\mp\qq})} 
   \end{cases} 
\end{align}

\subsection{Electron-impurity kinetic corrections}

In this case, $\Sigma[G]$ is given by Eq. \ref{eq:sigmaimp}. So the correction to the real part of the self-energy due to the external field is negligible: $\delta\real  \Sigma \sim 0 $, because $G^\textR$ and $G^\textA$ are taken to always be in equilibrium. This gives the following expressions for kinetic corrections (see derivations in Appendix Sec. \ref{appsec:kineticcorimp}):
\begin{align}
 \commutator{\real \Sigma^\texteq (\omega)}{\delta G^{\textK}(\omega)} &\rightarrow -\dfrac{1}{2}\left( \sum_{\pp'} \left( \left[(\delta h_{\pp}V_{\pp\pp'}) \odot  \Pi_{\pp\pp'}\right] + \delta h_{\pp}\left[V_{\pp\pp'} \odot  \Pi_{\pp\pp'}\right] \right) V_{\pp\pp'}^\dagger - h.c.\right), \\
 \commutator{\delta \SigmaK (\omega)}{\real G^{\texteq}(\omega)} &\rightarrow \dfrac{1}{2}\left( \sum_{\pp'} \left( \left[ (V_{\pp \pp'}\delta h_{\pp'}) \odot \Pi_{\pp\pp'} \right] + \left[ V_{\pp \pp'} \odot \Pi_{\pp\pp'} \right]\delta h_{\pp'}\right)V_{\pp\pp' }^\dagger - h.c. \right),
\end{align}
where $[\Pi_{\pp\pp'}]_{nn'}$ is defined as
\begin{align*}
    [\Pi_{\pp\pp'}]_{nn'} \equiv \begin{cases} 0,  \text{ if } \epsilon_{n'\pp'} = \epsilon_{n\pp } \\ 
   \dfrac{1}{\epsilon_{n\pp} - \epsilon_{n'\pp'} } 
   \end{cases} 
\end{align*}

To our knowledge, these kinetic correction terms have not been given in prior works and, therefore, constitute a novel extension of the standard theory. As shown by their matrix elements in Appendices \ref{appsec:kineticcor} and \ref{appsec:kineticcorimp}, the off-diagonal elements of these terms are non-vanishing even in the case of when coherences are explicitly ignored, i.e., when the $\delta h_{\pp}$ is assumed to be diagonal.

\section{Linearized matrix BTE} \label{sec:matrixBTE}
Here we collect the terms of the linearized electronic BTE and express them explicitly in terms of their matrix elements. All terms except the pure field-coupling are the sum of two contributions: one from electron-phonon coupling and the other from electron-impurity scattering.

\paragraph*{Pure field coupling:}

\begin{align}
\convolcommutator{G^{-1}_{0}}{G^{\textK}}_{nn'} \rightarrow 
    \delta h_{nn'\pp}\epsilon_{n'\pp} + h_{n\pp}^\texteq U_{nn'}  - \dfrac{i}{2}[\partial_{\RR}U(\RR)]_{nn'}\cdot\partial_{\pp}h_{\pp n'}^\texteq - h.c.
\end{align}

\paragraph*{Kinetic correction 1:}

\begin{align}
     &-\commutator{\real \Sigma^\texteq}{\delta G^{\textK}}_{nn'}^{\text{e-imp}}  \rightarrow \dfrac{1}{2}\sum_{\pp'ij}   \left( \delta h_{ni\pp}V_{ij\pp\pp'}V_{jn'\pp\pp'}^\dagger( \Pi_{nj\pp\pp'} + \Pi_{ij\pp\pp'} ) - h.c.\right), \\
    &-\commutator{\real \Sigma^{\texteq}}{\delta\GK}_{nn'}^{\text{e-ph}}   \rightarrow \dfrac{1}{4}\sum_{s\pp'ij}\left(\delta h_{ni\pp}g_{sij\pp\pp'}^\dagger g_{sjn'\pp\pp'}\left(\Piell_{s\pp\pp'nj} + \Piell_{s\pp\pp'ij} + h_{\pp'j}^{\texteq}(\bar{\Pi}_{s\pp\pp'nj} + \bar{\Pi}_{s\pp\pp'ij})\right) - h.c.\right) \\
    &-\commutator{\real \delta \Sigma}{G^{\textK, \texteq}}_{nn'}^{\text{e-ph}} \rightarrow\dfrac{1}{4}\sum_{s\pp'ij}\left( h_{\pp n}^{\texteq}g_{sni\pp\pp'}^\dagger g_{sjn'\pp\pp'}\delta h_{ij\pp'}(\bar{\Pi}_{s\pp\pp'ni}+\bar{\Pi}_{s\pp\pp'nj})- h.c.\right)
\end{align}

\paragraph*{Kinetic correction 2:}

\begin{align}
      & - \commutator{\delta \SigmaK }{\real G^{\texteq}}^{\text{e-imp}}_{nn'} \rightarrow  -\dfrac{1}{2}\sum_{\pp'ij}   \left( V_{ni\pp\pp'} V_{jn'\pp\pp'}^\dagger \delta h_{ij\pp'} (\Pi_{nj\pp\pp'} + \Pi_{ni\pp\pp'}) - h.c.\right), \\
      &-\commutator{\delta \SigmaK}{\real G^{\texteq}}^{\text{e-ph}}_{nn'} \rightarrow  -\dfrac{1}{4}\sum_{s\pp'ij} g_{sni\pp\pp'}^\dagger g_{sjn'\pp\pp'}\delta h_{ij\pp'}\Piell_{s\pp\pp'nj}  - h.c.
\end{align}

\paragraph*{In-scattering:}
\begin{align}
    &\dfrac{i}{2}\anticommutator{\delta \SigmaK_{\pp}}{A_{\pp}^{0,\texteq}}^{\text{e-imp}}_{nn'} \rightarrow \dfrac{i\pi}{2}\sum_{ij\pp'}\Big\{ V_{ni\pp\pp'}V^{\dagger}_{jn'\pp\pp'} \delta h_{\pp' ij}\left(\Delta_{\pp\pp'ni} + \Delta_{\pp\pp'nj}\right) + h. c.\Big\} \\
    &\dfrac{i}{2}\anticommutator{\delta \SigmaK_{\pp}}{A_{\pp}^{0,\texteq}}^{\text{e-ph}}_{nn'} \rightarrow \dfrac{i\pi}{4}\sum_{sij\pp'}\left\{ g^{\dagger}_{sni\pp\pp'}g_{sjn'\pp\pp'} \left[ \delta h_{\pp'ij}\left( \Deltaell_{\pp\pp'sni} + \Deltaell_{\pp\pp'snj} \right)  \right] + h.c. \right\}
\end{align}

\paragraph*{Out-scattering:}
\begin{align}
    -\dfrac{i}{2}\anticommutator{\Gamma^{0}_{\pp}}{\GK_{\pp}}_{nn'}^{\text{e-imp}} \rightarrow  -\dfrac{i\pi}{2}\sum_{ij\pp'}&\Big\{ V_{ij\pp\pp'}V^{\dagger}_{jn'\pp\pp'} \delta h_{\pp ni}\left(\Delta_{\pp\pp'nj} + \Delta_{\pp\pp'ij}\right) + h. c.\Big\},\\
    -\dfrac{i}{2}\anticommutator{\Gamma^{0}_{\pp}}{\GK_{\pp}}_{nn'}^{\text{e-ph}} \rightarrow -\dfrac{i\pi}{4}\sum_{sij\pp'}&\Big\{ g^{\dagger}_{sij\pp\pp'}g_{sjn'\pp\pp'}h^{\texteq}_{\pp'j}\delta h_{\pp ni}\left( \bar{\Delta}_{\pp\pp'snj} + \bar{\Delta}_{\pp\pp'sij } \right) \nonumber \\
    &+ g^{\dagger}_{sni\pp\pp'}g_{sjn'\pp\pp'}h^{\texteq}_{\pp n}\delta h_{\pp'ij}( \bar{\Delta}_{\pp\pp'sni} + \bar{\Delta}_{\pp\pp'snj} ) \nonumber \\ 
    &+ g^{\dagger}_{sij\pp\pp'}g_{sjn'\pp\pp'} \delta h_{\pp ni} \left( \Deltaell_{\pp\pp'snj} + \Deltaell_{\pp\pp'sij} \right) + h.c. \Big\}
\end{align}

In Appendix Sec. \ref{appsec:standardtheory}, we show that under appropriate approximations, the expressions above correspond to the standard electronic BTE.

Our general matrix BTE reveals multiple connections between the Wigner function matrix elements corresponding to different bands. Contributions to the coherence terms come not only from the occupations within the same band but also from the coherences with other bands. On the right hand side, the $\delta$-functions select only those scattering processes that follow the energy conservation law. This describes the redistribution of probability due to scattering processes. The left hand side shows that kinetic corrections to the probability distribution are proportional to the inverse energy difference between states of different bands. Interestingly, this form coincides with the first-order perturbation theory correction to the eigenstates of a perturbed system, which also involves contributions from all bands. Note that the kinetic corrections vanish only when the off-digonal elements of both the self-energies and the Wigner function are ignored. Our findings differ from those in Ref. \cite{cepellotti2021interband}, where the application of the RTA and the neglect of kinetic terms decouple the density matrix elements. In that simplified case, non-equilibrium Wigner function matrix elements do not influence one another.

\section{Steady state direct current and conductivity} \label{sec:conductivity}
In this section we derive the expression for the charge conductivity. We start from the \textit{ab initio} definition of the microscopic charge current \cite{RevModPhys.58.323, ponce2020first}
\begin{equation}
    \pmb{\mathscr{J}}^{\text{c}}(1, 2) = 2\dfrac{-e}{2m_{e}} \left(\partial_{\mathbf{r}_{1}} - \partial_{\mathbf{r}_{2}}\right)G^{<}(1, 2)\Big|_{1 = 2},
\end{equation}
where the multiplicative factor $2$ counts the spin degrees of freedom and $m_{e}$ is the electronic mass.

Expressing the above in the Wigner representation and in the eigenbasis of $H^{0}_{\pp}$, we get the following expression for the average macroscopic direct current:
\begin{equation}
    \langle \pmb{\mathscr{J}}^{\text{c}} \rangle \equiv \dfrac{1}{V}\int\dfrac{\text d\omega}{2\pi}\text{Tr}\sum_{\pp}\pmb{\mathscr{J}}^{\text{c}}_{\pp} = -ie\dfrac{1}{V}\text{Tr}\sum_{\pp}v_{\pp}\int\dfrac{\text d\omega}{2\pi}\GK_{\pp} = \dfrac{2e}{V}\text{Tr}\sum_{\pp}v_{\pp}f_{\pp},
\end{equation}
where we have identified the matrix elements of $\pp/m_{e}$ with the non-diagonal velocity tensor $v_{\pp}$. Above, $V$ is the volume of the crystal and the trace over the band space is denoted by Tr. Note that because the average current is a traced quantity, the ordering of $v_{\pp}$ and $f_{\pp}$ is unimportant. However, since the diagonal terms of the final expression contain contributions from the off-diagonal terms of both $v_{\pp}$ and $f_{\pp}$, the effect of the coherences is duly captured. This expression is also formally identical to the one in Ref. \cite{cepellotti2021interband}.

Next, we derive the expression for the conductivity. We first note that the contribution to the current from the equilibrium (Fermi-Dirac) part of $f_{\pp} = f^{\texteq}_{\pp} + \delta f_{\pp}$ trivially vanishes, yielding
\begin{equation}
    \langle \pmb{\mathscr{J}}^{\text{c}} \rangle = \dfrac{2e}{V}\sum_{\pp}\sum_{mn}\mathbf{v}_{mn\pp}\delta f_{nm\pp} = \dfrac{2e}{V}\sum_{\pp}\sum_{m}\mathbf{v}_{m\pp}\delta f_{m\pp} + \dfrac{2e}{V}\sum_{\pp}\sum_{m \neq n}\mathbf{v}_{mn\pp}\delta f_{nm\pp},
\end{equation}
where in the last line we have separated the expression into the diagonal and off-diagonal contributions. We notice that unlike in the Ref. \cite{cepellotti2021interband}, the diagonal part contains both the population and coherence effects since our coherence augmented BTE is beyond the RTA and mixes bands. The off-diagonal part, similarly, also contains both these effects.

We now operate within the linear response regime and write
\begin{equation}
    \delta f_{mn\pp} \doteq \beta \mathbf{X}_{mn\pp}\cdot\mathbf{E},
\end{equation}
where $\beta$ is the inverse temperature energy, and $\mathbf{X}$ is the (vectorial) response to the electric field.

Then, from Ohm's law $\langle \pmb{\mathscr{J}}^{\text{c}} \rangle = \sigma \mathbf{E}$, the conductivity Cartesian tensor is given by
\begin{equation}
    \sigma = \dfrac{2e\beta}{V}\sum_{\pp}\text{Tr}\left(v_{\pp}\times \mathbf{X}_{\pp}\right),
\end{equation}
where the tensor product ($\times$) is over the Cartesian space and the trace is over the band space.

This expression should be applicable under the conditions where the quasiparticle + weak-field approximation is valid. Note again that this conductivity expression includes both the population and interband coherence effects.

\section{Summary and future work}
In this work, we derive a new linearized electronic BTE for a homogeneous, weak electric field, \textit{ab initio}. Our transport equation considers the band non-diagonality of both the interacting self-energies and the Wigner function. As such, it takes a matrix form. We include the e-ph and e-imp interactions within Matthiesen's rule. We find that the kinetic corrections, i.e., the effects of interactions on the action of the driving field on the particles, are non-zero. These corrections scale inversely with the interband energy difference and sum contributions from every band of the Wigner function. We derive expressions for the collision side, which are far more general compared to the theories that ignore the off-diagonal terms of either the self-energies or the Wigner function. We provide our final working equation in a form that is amenable to direct numerical computation. We also show that the kinetic side of our theory corresponds to the one proposed in Ref. \cite{cepellotti2021interband} if the kinetic corrections are ignored. Our collision side, however, has a very different expression which stems from the fact that we do not make the diagonal approximation for the self-energies and make no relaxation time approximations.

We list here some of the possible research directions that we wish to venture into in the future:
\begin{itemize}
    \item \textbf{Numerical implementation} Our matrix BTE can be solved numerically to study the effect of the interband effects in real materials. This is a challenging task, especially with parameters-free methods that use, for example, density functional and related theories. While these parameters-free methods can provide all the necessary ingredients that constitute the matrix BTE, compared to the standard BTE, the numerical complexity is significantly higher due to the presence of several extra terms and also the general matrix form of these terms. We think that it would be prudent to focus first on low cost methods for the computation of the interactions vertices, similar to the one, for example, given in Ref. \cite{peng2025efficient}, before attempting a numerical solution of this equation.

    \item \textbf{Magnetotransport} Over the past years, the standard electronic BTE has been used to compute the magnetotransport properties in a parameters-free manner \cite{PhysRevB.98.201201, PhysRevB.103.L161103}. The methodology discussed in this work can be extended to include a magnetic field. Now, in order to study the interband effects on magnetotransport, a vector potential must be introduced in the driving term. The resulting transport equation, however, must be gauge invariant with respect to both the scalar and the vector potentials. This issue of gauge invariance in the context of Boltzmann-like equations has been discussed previously in Refs. \cite{RevModPhys.58.323} and \cite{mahan2013many}.

    \item \textbf{Temperature gradient} Considering the effect of a temperature gradient as a driving field is a difficult issue. This is because a temperature gradient does not exert any mechanical force on the system. In other words, there is no time dependent term that can be added to the Hamiltonian that would generate the temperature gradient as a field. In Ref. \cite{meier1969green}, Meier derives the phonon BTE using a Green's function approach. There, a time dependent external field term was added to the Hamiltonian, the role of which is to enable the use of the quantum kinetic formalism. In the steady state, however, the effect of this term has to be eliminated. The physical meaning of this term remains unclear to us. Interestingly, a gravitational field analogue of the generator of a temperature gradient was proposed by Luttinger in Ref. \cite{luttinger1964theory}.

    \item \textbf{Off-shell effects} In this work, we used the on-shell or quasiparticle approximation from the onset. The benefit of this is that we do not have any equation of motion connecting the retarded/advanced Green's function and the Keldysh Green's function that has to be solved self-consistently. Another benefit is that the quasiparticle BTE does not have any surviving dependence on the frequency variable. A possible step forward to include off-shell effects would be to retain the interacting system's spectral function, while ignoring the self-consistency loop between the retarded/advanced and Keldysh sectors. Essentially, here the key approximation would be that the equilibrium and nonequilibrium spectral functions would be the same. The final working equation would, of course, be one dimension higher in complexity compared to the one derived in this work.

    \item \textbf{Phonon matrix BTE} The methodology employed here can also be used for the case of phonons. Indeed, some of the terms we derived on the left hand side are completely analogous to the ones in Ref. \cite{PhysRevX.12.041011}. Here, too, arises the question of what term in the Hamiltonian drives the phonon system out of equilibrium. A second issue is connected to how a phonon heat current can be defined. This has been discussed extensively in Ref. \cite{ercole2016gauge}.

    \item \textbf{Dragful electron-phonon matrix BTEs} In this work, we made the dragless approximation which demands that the phonon Wigner function be taken to remain in equilibrium. However, in an interacting electron-phonon gas, driving one subsystem out of equilibrium naturally does the same to the other \cite{peierls1930theorie, gurevich1989electron}. Following the construction of the phonon matrix BTE, a possible research direction would be to derive the coupled electron-phonon matrix BTEs where the non-equilibrium states of both system's Wigner functions would be retained. This would significantly generalize the current state of the art dragful electron-phonon BTEs as implemented in Ref. \cite{protik2022elphbolt}.
\end{itemize}

\section{Appendices}
\subsection{Miscellaneous mathematical relations}
\paragraph*{Hadamard product} Here we give a few useful relations involving the Hadamard product. The first of these is
\begin{align}
    A \equiv \int\textd\omega [\delta_{\pp}] M [\delta_{\pp'}] = M \odot [\delta_{\pp\pp'}], 
\end{align} 
where $[\delta_{\pp}]$ is a (diagonal) matrix of $\delta(\omega - \varepsilon_{\pp}) \forall \pp$, such that $[\delta_{\pp}]_{nm} = \delta_{nm}\delta(\omega - \varepsilon_{n\pp})$, $n$ being the band index; $[\delta_{\pp \pp'}]_{nm} \equiv \delta(\varepsilon_{n\pp} - \varepsilon_{m\pp'})$; and $M$ is a square matrix in the band space.

The proof is as follows: 
\begin{align}
    A_{nm} = \sum_{ij}\int\textd\omega \delta_{ni}\delta(\omega - \varepsilon_{\pp,n})  M_{ij} \delta_{jm}\delta(\omega - \varepsilon_{\pp',m}) = M_{nm}\delta(\varepsilon_{\pp,n} - \varepsilon_{\pp',m})) = [M \odot [\delta_{\pp\pp'}]]_{nm}. 
\end{align}

We show further that the following is true for the multiplication of diagonal matrices $A$ and $C$ with a square matrix $B$: $ABC = B \odot \Delta_{AC}$, where $[\Delta_{AC}]_{ij} = A_iC_j$:
\begin{align}
    [ABC]_{ij} = \sum_{k,m}A_{ik}B_{km}C_{mj} = \sum_{k,m}A_{i}\delta_{ik} B_{km}C_{j}\delta_{mj} = B_{ij}A_iC_j.
\end{align}

We also list the following properties of the Hadamard products:
\begin{align}
    \left[A\odot (HG)\right] = \left[(HA)\odot G\right] = H\left[A\odot G\right] \nonumber \\
    \left[A\odot (GH)\right] = \left[(AH)\odot G\right] = \left[A\odot G\right]H,
\end{align}
where $H$ is a  diagonal matrix and $A$, $G$ are arbitrary  matrices. These follow straightforwardly by inspecting the matrix elements.

\paragraph*{Operator derivative} For any operator $B$, the derivative of its inverse, $B^{-1}(\pp)$, can be derived from the identity $I \equiv BB^{-1}$ in the follow manner:
\begin{align} \label{eq:derivinvop}
    \partial_{\pp}I = 0 = \partial_\pp (BB^{-1}) = (\partial_\pp B)B^{-1} + B\partial_\pp B^{-1}   \implies  \partial_\pp B^{-1} = -B^{-1}(\partial_\pp B)B^{-1}.
\end{align}

\paragraph*{Diagonal part of Wigner functions} The following identities are useful when converting expressions given in terms of $h$ and $l^{0}$ into those in terms of the usual $f$ and $n^{0}$. Note that these expressions are valid only for the diagonal components of these quantities.
\begin{align} \label{eq:wigneridentities}
    &1 + l^{0}_{s-\qq}h_{n'\pp'} - l^{0}_{s-\qq}h_{n\pp} - h_{n'\pp'}h_{n\pp} = 4f_{n\pp}(1 - f_{n'\pp'})(1 + n^{0}_{s-\qq}) - 4(1 - f_{n\pp})f_{n'\pp'}n^{0}_{s-\qq} \nonumber \\
    &-1 + l^{0}_{s\qq}h_{n'\pp'} - l^{0}_{s\qq}h_{n\pp} + h_{n'\pp'}h_{n\pp} = 4f_{n\pp}(1 - f_{n'\pp'})n^{0}_{s\qq} - 4(1 - f_{n\pp})f_{n'\pp'}(1 + n^{0}_{s\qq})
\end{align}

The above can be straightforwardly verified by direct computation using Eq. \ref{eq:wignerandoccup}.

\subsection{Commutation of frequency integration and basis change} \label{appsec:basischange}
Here we drop the common momentum index in all variables for brevity. Expressing everything in the eigenbasis of the non-interacting Hamiltonian means that we change the basis set using the transformation matrix $S: S^\dagger H^0S = \text{diag}(\epsilon_1, \epsilon_2, ...)$, where $S$ is the matrix constructed out of eigenvectors of $H^0$, $S = \left\{|\phi_1\rangle, |\phi_2 \rangle, ...\right\}$. Now, because $H^0$ does not depend on $\omega$, $S$ also does not depend on $\omega$. This implies that for an arbitrary matrix $A$, we have 
\begin{equation}
 A' \equiv \int \textd\omega S^\dagger A(\omega) S = S^\dagger \left[\int \textd\omega A(\omega)\right]S.
\end{equation}

In the main text, we use the same variable name after the basis change.

\subsection{One-body density matrix, Wigner function, and Keldysh Green's function}\label{appsec:wigner}

Here we give the relationships between the one-body density matrix, Wigner function, and the Keldysh Green's function. We also derive the relationship between the standard Bose/Fermi-like distributions functions with the Wigner function used in this work.

From definition, the one-body density matrix is $\rho_1(\mathbf{r}_1, \mathbf{r}_2, t) \equiv \text{Tr}[\rho \psi^\dagger(\mathbf{r}_1, t)\psi(\mathbf{r}_2, t)]$, where $\rho$ is the full density matrix of the many-body system. It is directly connected to the lesser Green's function $G^{<}(\mathbf{r}_1, t_1, \mathbf{r}_2, t_2) \equiv \mp i\text{Tr}[\rho \psi^\dagger(\mathbf{r}_1, t_1)\psi(\mathbf{r}_2, t_2)]$ \cite{RevModPhys.58.323} under equal time conditions. Here the upper (lower) sign corresponds to a Bose (Fermi) field. That is, $\rho_1(\mathbf{r}_1, \mathbf{r}_2, t) = \pm i G^<(\mathbf{r}_1, t, \mathbf{r}_2, t)$. The Wigner distribution function is defined as \cite{rammer2007quantum} 
\begin{equation}
     W(\RR, \pp, t) \equiv \int \textd \mathbf{r} e^{i\pp\cdot\mathbf{r}} \rho_1\left(\RR+\dfrac{\mathbf{r}}{2}, \RR-\dfrac{\mathbf{r}}{2}, t\right),
\end{equation}
where the Fourier transform is understood to act elementwise.

We will first demonstrate how the occupation function introduced in the \textit{Ansatz} is related to the Wigner distribution function. 

The Keldysh and the lesser Green's functions are related by \cite{RevModPhys.58.323}
\begin{equation} \label{eq:GlesserandGK}
     G^< = \dfrac{1}{2}\GK + \dfrac{i}{2}A
\end{equation}

We define a Wigner function
\begin{equation}
    g_{\pp}(\RR, T) \equiv -\int\dfrac{\textd \omega}{2\pi i}\GK
\end{equation}

Combining this with the relation Eq. \ref{eq:GlesserandGK}, we get
\begin{align*}
     g_\pp(\RR, T) = -\dfrac{1}{2\pi i} \int \textd\omega  2G^<(\RR, \pp, \omega, T) + \dfrac{1}{2\pi i} \int \textd\omega iA(\RR, \pp, \omega, T).
\end{align*}

The first integral gives
\begin{align*}
     -\dfrac{1}{2\pi i} 2\int  &\textd t\underbrace{\int \textd\omega e^{-i\omega t}}_{2\pi\delta(t)}\int \textd\mathbf{r} e^{i\pp\cdot\mathbf{r}}G^<\left(\RR + \dfrac{\mathbf{r}}{2}, T + \dfrac{t}{2},\RR - \dfrac{\mathbf{r}}{2},T - \dfrac{t}{2}\right) \\
     &= 2i \int \textd\mathbf{r}e^{i\pp \mathbf{r}}G^<\left(\RR + \dfrac{\mathbf{r}}{2}, T,\RR - \dfrac{\mathbf{r}}{2},T\right) \\
     &= \mp 2i \int \textd\mathbf{r} e^{i\pp\mathbf{r}} i\rho_1\left(\RR + \dfrac{\mathbf{r}}{2}, \RR - \dfrac{\mathbf{r}}{2}, T\right) \\
     &= \pm 2 W(\RR, \pp, T).
\end{align*}

And the second integral evaluates to the unit matrix $\mathbb{1}$. The proof is given in Appendix subsection. \ref{appsec:specfun}. 

As a result, we have
\begin{equation} \label{eq:wignerandoccup}
      g_\pp(\RR, T) = \mathbb{1} \pm 2W(\RR, \pp, T).
\end{equation}

In the literature, $W$ is denoted $n(f)$ for a Bosonic (Fermionic) system. In this work we call both $g$ and $W$ Wigner functions. In the main body of the paper, we used $g = l(h)$ for phonons (electrons). Note that in equilibrium, the above trivially leads to Eqs. \ref{eq:phononl} and \ref{eq:electronh}.

\subsection{Normalization of spectral function}\label{appsec:specfun}
Here we derive the normalization of the spectral function for the general, non-diagonal case.

We use the standard definitions for the retarded and advanced Green's functions and the spectral function
\begin{align}
    G^\text{R}_{\sigma_1\sigma_2}(\mathbf{r}_1, t_1, \mathbf{r}_2, t_2) &\equiv -i\theta(t_1 - t_2)\left\langle[\psi^\dagger_{\sigma_1}(\mathbf{r}_1, t_1), \psi_{\sigma_2}(\mathbf{r}_2, t_2)]_\mp\right\rangle \nonumber \\
    G^\text{A}_{\sigma_1\sigma_2}(\mathbf{r}_1, t_1, \mathbf{r}_2, t_2) &\equiv i\theta(t_2 - t_1)\left\langle[\psi^\dagger_{\sigma_1}(\mathbf{r}_1, t_1), \psi_{\sigma_2}(\mathbf{r}_2, t_2)]_\mp\right\rangle \nonumber \\
    A_{\sigma_1\sigma_2}(\mathbf{r}_1, t_1, \mathbf{r}_2, t_2) &\equiv i(G_{\sigma_{1}\sigma_{2}}^\text{R} - G_{\sigma_{1}\sigma_{2}}^\text{A}),
\end{align}
where $\sigma$ represents a general set of quantum numbers required to define a one-particle state. 

Then,
\begin{align*}
     \int\textd\omega  A_{\sigma_1\sigma_2}(\RR, \pp, \omega, T) &= \int \textd t\underbrace{\int \textd\omega e^{-i\omega t}}_{2\pi\delta(t)} \int\textd\mathbf{r} e^{i\pp\mathbf{r}} A_{\sigma_1\sigma_2}\left(\RR + \dfrac{\mathbf{r}}{2}, T+ \dfrac{t}{2}, \RR - \dfrac{\mathbf{r}}{2}, T - \dfrac{t}{2}\right)  \\
     &= 2\pi \int\textd\mathbf{r} e^{i\pp\mathbf{r}}  A_{\sigma_1\sigma_2}\left(\RR + \dfrac{\mathbf{r}}{2}, T, \RR - \dfrac{\mathbf{r}}{2}, T \right) \\
     &= 2\pi \int \textd\mathbf{r} e^{i\pp\mathbf{r}} \left\langle[\psi^\dagger_{\sigma_1}(\RR + \dfrac{\mathbf{r}}{2}, T), \psi_{\sigma_2}(\RR - \dfrac{\mathbf{r}}{2},T)]_\mp\right\rangle,
\end{align*}

The expectation value in the last expression yields
\begin{align*}
     \left\langle[\psi^\dagger_{\sigma_1}\left(\RR + \dfrac{\mathbf{r}}{2}, T\right), \psi_{\sigma_2}\left(\RR - \dfrac{\mathbf{r}}{2},T\right)]_\mp\right\rangle &= \text{Tr}\left[ \rho \left[U^{\dagger}\psi^\dagger_{\sigma_1}\left(\RR + \dfrac{\mathbf{r}}{2}\right) U, U^{\dagger}\psi_{\sigma_2}\left(\RR - \dfrac{\mathbf{r}}{2}\right) U\right]_\mp \right] \\
    &= \text{Tr}\left[\rho U^{\dagger}\left[ \psi^\dagger_{\sigma_1}\left(\RR + \dfrac{\mathbf{r}}{2}\right), \psi_{\sigma_2}\left(\RR - \dfrac{\mathbf{r}}{2}\right)\right]_\mp U\right] \\ 
    &= \text{Tr}\left[\rho U^{\dagger} \delta_{\sigma_1 \sigma_2} \delta(\mathbf{r})U\right] \\
    &= \delta_{\sigma_1 \sigma_2}\delta(\mathbf{r}),
\end{align*}
where $U \equiv \exp\left[-iHT\right]$ is the time evolution operator.

It then follows that
\begin{align*}
     \int \textd \omega A_{\sigma_1 \sigma_2}(\RR, \pp, \omega, T) = 2\pi \int \textd \mathbf{r} e^{i\pp\cdot\mathbf{r}} \delta(\mathbf{r}) \delta_{\sigma_1 \sigma_2}  = 2\pi \delta_{\sigma_1 \sigma_2}.
\end{align*}  

\subsection{Detailed balance}\label{appsec:detbalance}
Here we derive the detailed balance condition. This has already been shown by Rammer in Ref. \cite{rammer2007quantum} for the single-band case. Here we prove that the relation holds for the multiband case. Consider first the QKE in equilibrium, i.e., $U = 0$:
\begin{equation}\label{eq:QKE_eq}
    \convolcommutator{(G^{0, \texteq})^{-1}}{G^{\textK, \texteq}} - \convolcommutator{\real\Sigma^{\texteq}}{G^{\textK, \texteq}} - \convolcommutator{\Sigma^{\textK, \texteq}}{\real G^{\texteq}} = \dfrac{i}{2}\convolanticommutator{\Sigma^{\textK, \texteq}}{A^{\texteq}} - \dfrac{i}{2}\convolanticommutator{\Gamma^{\texteq}}{G^{\textK, \texteq}}.
\end{equation}

The aim is to prove that both sides of the equation vanish in equilibrium. To this end, we recall that in equilibrium, correlation functions are time-translationally invariant; that is, they depend only on the difference between time arguments, and also are translationally invariant. We also consider here a homogeneous system. As such, in the Wigner representation, these functions have no dependence on $T$ or $\RR$. As a result, in the Moyal product, all terms except the first one vanish. Thus, Eq. \ref{eq:QKE_eq} simplifies to the following form:
\begin{align}
    \commutator{G^{-1, \texteq}_{0}(\pp, \omega)}{G^{\textK, \texteq}(\pp, \omega)} - &\commutator{\real\Sigma^{\texteq}(\pp, \omega)}{G^{\textK, \texteq}(\pp, \omega)} - \commutator{\Sigma^{\textK,\texteq}(\pp, \omega)}{\real G^{\texteq}(\pp, \omega)} \nonumber \\
    &= \dfrac{i}{2}\anticommutator{\Sigma^{\textK, \texteq}(\pp, \omega)}{A^{\texteq}(\pp, \omega)} - \dfrac{i}{2}\anticommutator{\Gamma^{\texteq}(\pp, \omega)}{G^{\textK,\texteq}(\pp, \omega)}
\end{align}

Also, we will use the diagonal elements of the Dyson equation \cite{rammer2007quantum}, which in equilibrium look as follows: 
\begin{equation}
    \commutator{G^{-1}_0 - \Sigma^{\textR/\textA}}{G^{\textR/\textA}} =  0 
\end{equation}

The fluctuation-dissipation theorem states the following relations: 
\begin{align}\label{FDT1}
    G^{\textK, \texteq}(\pp, \omega) = h_0(\omega)\left(G^{\textR,\texteq}(\pp, \omega) - G^{\textA,\texteq} (\pp, \omega) \right), \\
    \Sigma^{\textK, \texteq}(\pp, \omega) = h_0(\omega)\left(\Sigma^{\text{R},\texteq}(\pp, \omega) - \Sigma^{\text{A},\texteq}(\pp, \omega) \right),
\end{align}
where $h_0 = \text{tanh}\left( \dfrac{\omega\beta}{2}\right)$. 

Using these relations, the second and third terms in the LHS of the QKE become:
\begin{align*}
       \commutator{\real\Sigma^{\texteq}}{G^{\textK,\texteq}} + \commutator{\Sigma^{\textK,\texteq}}{\real G^{\texteq}} &=   h_0\left(\commutator{\Sigma^{\textR, \texteq} + \Sigma^{\textA, \texteq}}{G^{\textR, \texteq} - G^{\textA. \texteq}}   + \commutator{\Sigma^{\textR, \texteq} - \Sigma^{\textA, \texteq}}{G^{\textR, \texteq} + G^{\textA. \texteq}}\right) \\
    & =    h_0 \left( \commutator{\Sigma^{\textR, \texteq}}{G^{\textR,\texteq}} - \commutator{\Sigma^{\textA, \texteq}}{G^{\textA, \texteq}}\right)  
\end{align*}

Applying diagonal elements of the Dyson equations, the LHS: 
\begin{align*}
     \commutator{G^{-1}_0}{G^{\textK}} - \commutator{\real\Sigma}{G^{\textK}} - \commutator{\Sigma^{\textK}}{\real G} &=  h_0 \commutator{G^{-1}_0}{G^R - G^A} -h_0 \left( \commutator{\Sigma^\textR}{G^\textR} - \commutator{\Sigma^\textA}{G^\textA}\right) \\
    & =  h_0 (\commutator{G^{-1}_0 - \Sigma^{\textR}}{G^{\textR}} - \commutator{G^{-1}_0 - \Sigma^{\textA}}{G^{\textA}}) \\
    &= 0 
\end{align*}

So the detailed balance condition is as follows:
\begin{equation}
    \convolanticommutator{\Sigma^{\textK, \texteq}}{A^{\texteq}} - \convolanticommutator{\Gamma^{\texteq}}{G^{\textK, \texteq}} = 0.
\end{equation}
In other words, in- and out-scattering rates perfectly balance each other, keeping the occupations of the states to their equilibrium Fermi-Dirac values. Note that in equilibrium all quantities are band diagonal, i.e., there is no interband coherence/tunneling. Also note that the above proof does not make any assumption of diagonality of the self-energies.

\subsection{Detailed derivation of e-ph collision terms}\label{appsec:collision_eph}

\paragraph{In-scattering term:}

From the definition of the self-energy in the case of el-ph coupling, $\SigmaK$ is given by the following expression: 
\begin{align}
    \SigmaK = \dfrac{i}{2}\sum_{s\pp'}\dfrac{2}{\omega_{s\qq}} (\Lambda^{\text{R}} - \Lambda^{\text{A}}) \star (D^{\text{R}} - D^{\text{A}})+ \Lambda^{\text{K}}\star D^{\text{K}}   
\end{align}

After substituting $\Lambda^{\text{R}} - \Lambda^{\text{A}} = g^\dagger(G^{\text{R}} - G^{\text{A}})g = -ig^\dagger A^{0, \texteq}g $ and  expressions for $D^{\text{R}} - D^{\text{A}}$ and $D^{\text{K}}$, $\SigmaK$ is 
\begin{align*}
    \SigmaK &=\dfrac{i}{2}\sum_{s\pp'}\dfrac{2}{2\pi\omega_{s\qq}} \int \textd \omega' \left( -g^\dagger_sA_{\pp'}^{0, \texteq}(\omega')g_sB^{\texteq, s}_{0}(\omega - \omega', \qq) - i g^\dagger_s \GK_{\pp'}(\omega') g_s B^{\texteq, s}_{0}(\omega - \omega', \qq)l^{0, \texteq}_\qq(\omega - \omega')\right) \\ 
    &= -i\sum_{s\pp'}\dfrac{1}{2\pi\omega_{s\qq}} \int \textd \omega'g^\dagger_s\left[ A_{\pp'}^{0, \texteq}(\omega') + \dfrac{1}{2}\anticommutator{A_{\pp'}^{0, \texteq}(\omega')}{h_{\pp'}}l^{0, \texteq}_\qq(\omega - \omega') \right]g B^{\texteq, s}_{0}(\omega - \omega', \qq)   
\end{align*}

The integral term has a typical form for all terms of the formula above: 
\begin{align*}
    \int \dfrac{\textd\omega'}{2\pi\omega_{s\qq}} A_{\pp'}^{0, \texteq} B^{\texteq, s}_{0} l^{0, \texteq}_\qq &=  \pi \int \textd\omega' \delta(\omega' - H_{\pp'}^0)l^{0, \texteq}_\qq(\omega - \omega') [ \delta(\omega_{s\qq} - \omega + \omega') - \delta(\omega_{s\qq} + \omega - \omega')]  \\ &= \pi  \left( A_{\pp'}^+l_\qq^+  -  A_{\pp'}^-  l_\qq^-  \right), 
\end{align*}
where $A_{\pp'}^\pm = \delta(\omega - (H_{\pp'}^0 \pm \omega_{s\qq}))$, $l_\qq^\pm = l^{0, \texteq}_\qq(\pm\omega_{s\qq})$

As a result,
\begin{align}
    \SigmaK = - \dfrac{i\pi}{2}\sum_{s\pp'} g_s^\dagger\left[\anticommutator{A_{\pp'}^+l^+_\qq -  A_{\pp'}^-l^-_\qq}{h_{\pp'}} + 2(A_{\pp'}^+ - A_{\pp'}^-) \right]g_s = - \Sigma^{\textK \dagger} 
\end{align}

From the Hermitian property of the spectral function and the anti-Hermitian property of $\SigmaK$ it follows that
\begin{align}
    \dfrac{i}{2}\int \dfrac{\textd \omega}{-2\pi i} \anticommutator{\SigmaK}{A^{0,\texteq}} = \dfrac{i\pi}{4}(I^{(1)} + h.c.),
\end{align}
where
\begin{align}
    I^{(1)} = \sum_{s\pp'}\int \textd\omega \dfrac{A_\pp^{0,\texteq}}{2\pi} g_s^\dagger\left[\anticommutator{A_{\pp'}^+l^+_\qq -  A_{\pp'}^-l^-_\qq}{h_{\pp'}} + 2(A_{\pp'}^+ - A_{\pp'}^-) \right]g_s   
\end{align}

Let's take the integral for one of the terms:
\begin{align*}
    \dfrac{1}{2\pi}\int \textd\omega \left[ A_\pp^{0,\texteq}g_s^\dagger A_{\pp'}^+l^+_\qq h_{\pp'}g_s \right]_{nn'} &= \sum_{i}\int \textd\omega \delta(\omega - \epsilon_{ n\pp}) g^\dagger_{sni}\delta(\omega-(\epsilon_{i\pp'} +\omega_{s\qq}))l^+_\qq (h_{\pp'}g_{s})_{in'} \\
    &= \sum_{i} g^\dagger_{sni}\delta(\epsilon_{n\pp}-(\epsilon_{i\pp'} +\omega_{s-\qq}))l^+_{-\qq} (h_{\pp'}g_{s})_{in'} \\
    & = \left[\left[g_s^\dagger \odot (\Delta^{\text{emi}, s-\qq}_{\pp\pp'}l^+_{-\qq})\right]h_{\pp'}g_s\right]_{nn'}
\end{align*}

Here we changed the sign for $\qq$ in $l^+$ because energy and momentum conservation rules for the emission process give $\qq = \pp - \pp'$, while it was introduced as $\qq = \pp' - \pp$ originally.  

Calculating the other terms in a similar manner, finally, we obtain,
\begin{align}
    I^{(1)} &=  \sum_{s\pp'}\left[ g_s^\dagger \odot \Deltaell^{s}_{\pp\pp'}\right]h_{\pp'}g_s + \left[ (g_s^\dagger h_{\pp'}) \odot \Deltaell^{s}_{\pp\pp'}\right]g_s + 2 \left[ g_s^\dagger \odot \bar{\Delta}^{s}_{\pp\pp'}\right]g_s
\end{align} 

\paragraph{Out-scattering term:}
First we evaluate
\begin{align*}
    \Gamma = i(\Sigma^\textR - \Sigma^\textA) &= i\sum_{s\pp'}\dfrac{i}{\omega_{s\qq}}(\Lambda^\textK\star D^\textR + \Lambda^\textR\star D^\textK - \Lambda^\textA \star D^\textK - \Lambda^\textK\star D^\textA) \\
    & = - \sum_{s\pp'}\dfrac{1}{\omega_{s\qq}}(\Lambda^\textK\star(D^\textR - D^\textA)+ (\Lambda^\textR - \Lambda^\textA)\star D^\textK) \\
    &= - \sum_{s\pp'}\dfrac{1}{\omega_{s\qq}}\left( g^\dagger_s \left[-\dfrac{i}{2} \anticommutator{A_{\pp'}^{0, \texteq}}{h_{\pp'}}\right]g_s \star(-iB^{\texteq, s}_{0}) + g^\dagger_s(-iA_{\pp'}^{0, \texteq})g_s \star (-iB^{\texteq, s}_{0}l_{s\qq}^{0, \texteq}) \right)\\
    &=  \sum_{s\pp'}\dfrac{1}{\omega_{s\qq}}\left( \dfrac{1}{2}g^\dagger_s \anticommutator{A_{\pp'}^{0, \texteq}}{h_{\pp'}}g_s \star B^{\texteq, s}_{0} + g^\dagger_sA_{\pp'}^{0, \texteq}g_s \star B^{\texteq, s}_{0}l_{s\qq}^{0, \texteq}\right)
\end{align*}

Integrating similarly to $\Sigma^\textK$ case above, we get
\begin{align}
    \Gamma_\pp =  \dfrac{\pi}{2}\sum_{s\pp'}g^\dagger_s\left(\anticommutator{A_{\pp'}^+ - A_{\pp'}^-}{h_{\pp'}} + 2(A_{\pp'}^+l_\qq^+ - A_{\pp'}^- l_\qq^-)\right)g_s = \Gamma_\pp^\dagger
\end{align}

Then, as we did with the in-scattering term, using the Hermiticity of $\Gamma$ and the anti-Hermiticity of $\GK$, we find
\begin{align}
    \dfrac{i}{2}\anticommutator{\Gamma}{\GK} &\rightarrow \dfrac{i\pi}{8} \sum_{s\pp'} \biggl(  h_\pp \Bigl( \left[ g^\dagger_s\odot \bar{\Delta}^{s}_{\pp\pp'} \right]h_{\pp'} + \left[ (g_s^\dagger h_{\pp'})\odot \bar{\Delta}^{s}_{\pp\pp'} \right]  +2 \left[ g^\dagger_s\odot \Deltaell^{s}_{\pp\pp'} \right]\Bigl)g_s \nonumber \\
    & + \Bigl(\left[(h_{\pp}g_s^\dagger ) \odot \bar{\Delta}^{s}_{\pp\pp'}\right]h_{\pp'} + \left[ (h_{\pp}g_s^\dagger h_{\pp'})\odot \bar{\Delta}^{s}_{\pp\pp'} \right] +2 \left[ (h_{\pp}g^\dagger_s)\odot \Deltaell^{s}_{\pp\pp'}\right]\Bigl)g_s + h.c. \biggl)
\end{align}

\subsection{Detailed derivation of e-imp collision terms}\label{appsec:collision_eimp}

In this case $\Sigma[G]$ is given by Eq. \ref{eq:sigmaimp}. Taking into account that $V_{\pp' \pp}^T = V_{\pp\pp'}^\dagger$, we see that $\Sigma_\pp^{\textK \dagger} = - \SigmaK_\pp$ and $\Gamma^{0}_\pp = \sum_{\pp'} V_{\pp\pp'}A^{0, \texteq}_{\pp'}V_{\pp\pp'}^\dagger = \Gamma^{0\dagger}_\pp$. We will use these facts below.

\paragraph{In-scattering term:}

\begin{align*}
    \int \dfrac{\textd\omega}{-2\pi i}\anticommutator{\SigmaK_\pp(\omega)}{A_\pp^{0, \texteq}(\omega)} & = \sum_{\pp'}\int \dfrac{\textd\omega}{4\pi}\anticommutator{V_{\pp\pp'}\anticommutator{A_{\pp'}^{0, \texteq}}{h_{\pp'}}V_{\pp\pp'}^\dagger}{A_\pp^{0, \texteq}(\omega)} \\
    & =  \sum_{\pp'}\left(\int \dfrac{\textd\omega}{4\pi} A_\pp^{0, \texteq} V_{\pp\pp'} \anticommutator{A_{\pp'}^{0, \texteq}}{h_{\pp'}}V_{\pp\pp'}^\dagger  + h.c. \right)
\end{align*}

Taking the integral for one of the terms, we get
\begin{align*}
    \dfrac{1}{4\pi}\int \textd\omega \left[A_\pp^{0, \texteq} V_{\pp\pp'} A_{\pp'}^{0, \texteq}h_{\pp'}V_{\pp\pp'} \right]_{nn'} &= \pi \sum_{i}\int \textd\omega \delta(\omega - \epsilon_{ n\pp}) V_{ni}\delta(\omega-\epsilon_{i\pp'})(h_{\pp'}V^\dagger)_{in'} \\
    &= \pi \sum_{i} V_{ni}\delta(\epsilon_{n\pp}-\epsilon_{i\pp'}) (h_{\pp'}V^\dagger)_{in'} \\
    & = \pi\left[\left[V_{\pp\pp'} \odot \Delta_{\pp\pp'}\right](h_{\pp'}V_{\pp\pp'}^\dagger)\right]_{nn'}
\end{align*}

The other term can be evaluated similarly. Putting them together, we get
\begin{align*}
     \dfrac{i}{2}\anticommutator{\SigmaK_\pp(\omega)}{A_\pp^{0, \texteq}(\omega)} & \rightarrow  \dfrac{i\pi }{2} \sum_{\pp'}\left( \left[V_{\pp\pp'} \odot \Delta_{\pp\pp'}\right](h_{\pp'}V_{\pp\pp'}^\dagger) + \left[(V_{\pp\pp'}h_{\pp'}) \odot \Delta_{\pp\pp'}\right]V_{\pp\pp'}^\dagger  + h.c. \right)
\end{align*}

\paragraph{Out-scattering term:}
Similarly, we can show that
\begin{align*}
     \int \dfrac{d\omega}{-2\pi i}\anticommutator{\Gamma^{0}_\pp(\omega)}{\GK_\pp(\omega)} & = \sum_{\pp'}\int \dfrac{\textd\omega}{4\pi} \anticommutator{\Gamma^{0}_\pp(\omega)}{\anticommutator{h_{\pp}}{A_{\pp}^{0,\texteq}}} \\
    & =  \sum_{\pp'}\left(\int \dfrac{\textd\omega}{4\pi} \anticommutator{A_{\pp}^{0, \texteq}}{h_{\pp}}V_{\pp\pp'} A_{\pp'}^{0, \texteq}V_{\pp\pp'}^\dagger   + h.c. \right)
\end{align*}

Finally, we get
\begin{align*}
     -\dfrac{i}{2}\anticommutator{\Gamma^{0}_\pp(\omega)}{\GK_\pp(\omega)}& \rightarrow   -\dfrac{i\pi }{2} \sum_{\pp'}\left(\left[\Delta_{\pp\pp'} \odot (h_{\pp}V_{\pp\pp'})\right]V_{\pp\pp'}^\dagger + h_{\pp}\left[\Delta_{\pp\pp'} \odot V_{\pp\pp'}\right]V_{\pp\pp'}^\dagger + h.c. \right)
\end{align*}

\subsection{Detailed derivation of direct field coupling}\label{appsec:directfield}
Here we derive the direct field coupling term in the RHS of Eq. \ref{eq:qke}. Following the procedure outlined in the main text, we first consider combinations of the equilibrium correlation function and non-equilibrium deviations, giving first-order corrections to the Wigner function:
\begin{align}
    \convolcommutator{G^{-1}_{0}}{G^{\textK}} &\approx  \underbrace{\commutator{ G^{-1, \texteq}_{0}}{\delta G^{\textK}}}_{i} + \underbrace{\dfrac{i}{2}\commutator{G^{-1, \texteq}_{0}}{\delta G^{\textK}}_{p}}_{ii} - \underbrace{\dfrac{i}{2}\commutator{\delta G^{\textK}}{G^{-1, \texteq}_{0}}_{p}}_{iii}  \nonumber \\
    & + \underbrace{\commutator{\delta G^{-1}_{0}}{G^{\textK, \texteq}}}_{iv} + \underbrace{\dfrac{i}{2}\commutator{\delta G^{-1}_{0}}{G^{\textK, \texteq}}_{p}}_{v} - \underbrace{\dfrac{i}{2}\commutator{G^{\textK, \texteq}}{\delta G^{-1}_{0}}_{p}}_{vi}
\end{align}

\paragraph{Term $(i)$:}
\begin{align*}
    \commutator{\invGnot}{\delta G^{\textK}} = \cancelto{0}{\commutator{\omega}{\delta \GK}} - \commutator{H^{0}_{\pp} }{\delta\GK} \rightarrow \commutator{\delta h_{\pp}}{\varepsilon_{\pp}},
\end{align*}

\paragraph{Terms $(ii)-(iii)$:}
\begin{align*}
    \dfrac{i}{2}\commutator{\invGnot}{\delta \GK}_{p} &=  \dfrac{i}{2}\left(-\partial_{\RR}U(\RR)\cdot\partial_{\pp}\delta \GK + \partial_{\pp}H^{0}_{\pp}\cdot\partial_{\RR}\delta \GK \right)  
\end{align*}

Here $\partial_{\RR}\delta \GK = -\dfrac{i}{2}\anticommutator{A^\texteq}{\partial_\RR \delta h_\pp}$. But $ \delta G^\textK \sim \delta h_\pp\sim \partial_\RR U(\RR)$, so $ \commutator{\invGnot}{\delta \GK}_{p} = -\dfrac{i}{2}v_{\pp}\cdot\anticommutator{A^0}{\partial_\RR \delta h_\pp} + O((\partial_\RR U)^2)$. Similarly, $\commutator{\delta \GK}{\invGnot}_{p} = +\dfrac{i}{2}v_{\pp}\cdot\anticommutator{A^0}{\partial_\RR \delta h_\pp} +  O((\partial_\RR U)^2)$. As a result,
\begin{align*}
    \dfrac{i}{2}\commutator{\invGnot}{\delta \GK}_{p} -     \dfrac{i}{2}\commutator{\delta \GK}{\invGnot}_{p}   \rightarrow \dfrac{i}{2}\anticommutator{v_\pp}{\partial_\RR \delta h_\pp} 
\end{align*}

\paragraph{Term $(iv)$:}
\begin{align*}
    \commutator{\delta G^{-1}_{0}}{G^{\textK, \texteq}}  = [-U(\RR), -\dfrac{i}{2}\anticommutator{h_\pp^{\texteq}}{A^\texteq}] \rightarrow [h_\pp^\texteq, U(\RR)]
\end{align*}

\paragraph{Terms $(v) - (vi)$:}
\begin{align*}
   \dfrac{i}{2}\commutator{\delta G^{-1}_{0}}{G^{\textK, \texteq}}_{p}  &= \dfrac{i}{2}[-U(\RR), G^{\textK, \texteq}]_{p} = \dfrac{i}{2}\left( -\partial_{\RR}U(\RR)\cdot\partial_{\pp} G^{\textK, \texteq} + \cancelto{0}{\partial_{\pp}U(\RR)\cdot\partial_{\RR} G^{\textK, \texteq}}\right)  \\
   & = - \dfrac{i}{2}\partial_{\RR}U(\RR) \cdot \left( -i[\partial_\pp h_\pp^\texteq] A^{0,\texteq} -ih_\pp^\texteq [\partial_\pp A^{0,\texteq} ]\right) \\
   &\rightarrow -\dfrac{i}{2}\partial_{\RR}U(\RR)\partial_\pp h_\pp^\texteq - \dfrac{i}{2}\int \dfrac{ \textd\omega}{2\pi} h_\pp^\texteq [\partial_\pp A^{0,\texteq} ]
\end{align*}

Using the expression for the derivative of inverse of operators (Eq. \ref{eq:derivinvop}), we get
\begin{align}
    \partial_{\pp}A^{0,\texteq}_\pp = 2\imag[G_\pp^{\texteq,R}\partial_\mathbf{p}H_\pp^0G_\pp^{\texteq,R}] = -\real G_\pp^{\texteq, \mathrm{R}}\partial_\mathbf{p}H_\pp^0A^{0,\texteq}_\pp - A^{0,\texteq}_\pp\partial_\mathbf{p}H_\pp^0\real G_\pp^{\texteq, \mathrm{R}}
\end{align}

We now consider the integral $\int \dfrac{ \textd\omega}{2\pi} h_\pp^\texteq [\partial_\pp A^{0,\texteq}_\pp]$  for each matrix element:
\begin{align*}
   & \left[ \int \dfrac{ \textd\omega}{2\pi} h_\pp^\texteq [\partial_\pp A^{0,\texteq}_\pp]\right]_{ij} =  -\sum_{k}\int\dfrac{ \textd\omega}{2\pi} [h_\pp^\texteq]_{ik}  [\real G_\pp^{\texteq, \mathrm{R}}v_\pp A^{0,\texteq}_\pp + A^{0,\texteq}_\pp v_\pp \real G_\pp^{\texteq, \mathrm{R}}]_{kj}
\end{align*}

Here, the real part of the retarded Green's function in the eigenbasis of $H_{\pp}^0$ is  $[\real G_\pp^{\texteq, \mathrm{R}}]_{ij}(\omega) = \mint{-} \textd \omega' \dfrac{1}{\omega - \omega'}\delta(\omega' - \epsilon_{i\pp})\delta_{ij}$ and $\mint{-}$ denotes the principal value of the integral. In the integral we are considering, only $\real G_\pp^{\texteq, \mathrm{R}}$ and $A^{0,\texteq}_\pp$ depend on $\omega$. Therefore, 
\begin{align*}
    \sum_{ml}\int \dfrac{ \textd\omega}{2\pi} [\real G_\pp^{\texteq, \mathrm{R}}]_{km}v_{\pp, ml}[A^{0,\texteq}_\pp]_{lj} & =\sum_{ml} \int \dfrac{ \textd\omega}{2\pi} \mint{-} \textd \omega' \dfrac{1}{\omega - \omega'}\delta(\omega' - \epsilon_{k\pp})\delta_{km}v_{\pp, ml}2\pi\delta(\omega - \epsilon_{j\pp})\delta_{lj}  \\ 
   & = v_{\pp, kj} \mint{-} \textd \omega'\delta(\omega' - \epsilon_{k\pp})\int\textd\omega\dfrac{1}{\omega - \omega'}\delta(\omega - \epsilon_{j\pp}) \\
   & = v_{\pp, kj}\mint{-} \textd \omega'\dfrac{\delta(\omega' - \epsilon_{k\pp})}{\epsilon_{j\pp} - \omega '} \\
   &= \begin{cases} 0,  \text{ if } \epsilon_{j\pp} = \epsilon_{k\pp} \\ 
   \dfrac{v_{\pp, kj}}{\epsilon_{j\pp} - \epsilon_{k\pp}}, \text{otherwise} 
   \end{cases} 
\end{align*}

Using the same reasoning, we can show that
\begin{align*}
   & [A^{0,\texteq}_\pp v_\pp \real G^{\texteq, \mathrm{R}}]_{kj} = - [\real G^{\texteq, \mathrm{R}}A^{0,\texteq}_\pp v_\pp ]_{kj} 
\end{align*}

This implies that $\dfrac{i}{2}\int \dfrac{ \textd\omega}{2\pi} h_\pp^\texteq [\partial_\pp A^{0,\texteq}_\pp] = 0$. As a result, terms $(v) - (vi)$ becomes
\begin{align*}
   \dfrac{i}{2}\commutator{\delta G^{-1}_{0}}{G^{\textK, \texteq}}_{p} \rightarrow - \dfrac{i}{2}\anticommutator{\partial_{\RR}U(\RR)}{\partial_{\pp}h_{\pp}^\texteq}
\end{align*}

Putting these all together, we get
\begin{align}
    \convolcommutator{G^{-1}_{0}}{G^{\textK}} & \rightarrow \commutator{\delta h_{\pp}}{\varepsilon_{\pp}} + \dfrac{i}{2}\anticommutator{v_\pp}{\partial_\RR \delta h_\pp}  + [h_\pp^\texteq, U(\RR)] - \dfrac{i}{2}\anticommutator{\partial_{\RR}U(\RR)}{\partial_{\pp}h_{\pp}^\texteq}
\end{align}

\subsection{Detailed derivation of e-ph kinetic corrections}\label{appsec:kineticcor}

Here, $\real \Sigma$ is given by Eq. \ref{eq:real_sigma_eph} and $\SigmaK$ corresponds to the Keldysh component of the self-energy from Eq. \ref{eq:sigmaeph}. We first explicitly express the following quantities included there:
\begin{align*}
    & D_0^{\text{R},\texteq, s}(\omega, \qq) + D_0^{\text{A},\texteq, s}(\omega, \qq) = \omega_{s\qq}\left(\Pr\left[ \dfrac{1}{\omega - \omega_{s\qq} }\right] - \Pr\left[ \dfrac{1}{\omega + \omega_{s\qq} }\right]\right)\\
    & \Lambda^{\text{R},\texteq,s}(\omega, \pp, \pp')+\Lambda^{\text{A},\texteq, s}(\omega, \pp, \pp') = 2g_s^\dagger(\pp, \pp')\Pr\left[\dfrac{1}{\omega - H^0_{\pp'}}\right]g_s(\pp, \pp')\\ 
    & \Lambda^{\textK, s} (\omega, \pp, \pp') = \underbrace{-ig_s^\dagger(\pp, \pp')A^{0, \texteq}_{\pp'}(\omega)h_{\pp'}^{\texteq}g_s(\pp, \pp')}_{\Lambda^{\textK, \texteq, s}} \underbrace{- \dfrac{i}{2}g_s^\dagger(\pp, \pp')\anticommutator{A^{0, \texteq}_{\pp'}(\omega)}{\delta h_{\pp'}}g_s(\pp, \pp')}_{\delta\Lambda^{\textK, s}} 
\end{align*}

As a result,
\begin{align*}
    & \real \Sigma^{\texteq} = \dfrac{i}{2}\sum_{s\pp'}\dfrac{1}{\omega_{s\qq}} (\Lambda^{\textK, \texteq, s}\star(D_0^{\text{R},\texteq, s} + D_0^{\text{A},\texteq, s}) + (\Lambda^{\text{R},\texteq,s}+\Lambda^{\text{A},\texteq,s})\star D_0^{\text{K}, \texteq,s}) \\ 
    &\real\delta\Sigma=  \dfrac{i}{2}\sum_{s\pp'}\dfrac{1}{\omega_{s\qq}} \delta\Lambda^{\textK, s}\star(D_0^{\text{R},\texteq, s} + D_0^{\text{A},\texteq, s})
\end{align*}

Integrating over $\omega'$, we get
\begin{align}
    & \real \Sigma^{\texteq} = \dfrac{1}{2}\sum_{s\pp'}g^\dagger_s \left(P_{\pp'}^{+}(l^{0, \texteq}_\qq(\omega_{s\qq}) + h_{\pp'}^{\texteq}) -   P_{\pp'}^{-}(l^{0, \texteq}_\qq(-\omega_{s\qq}) + h_{\pp'}^{\texteq})\right)g_s =  \real \Sigma^{\texteq \dagger} \\ 
    &\real\delta\Sigma=   \dfrac{1}{4}\sum_{s\pp'}g^\dagger_s  \anticommutator{P_{\pp'}^{+} - P_{\pp'}^{-}}{\delta h_{\pp'}}g_s
\end{align}
where $P_{\pp'}^{+} \equiv \Pr\left[\dfrac{1}{\omega - (H_{\pp'}^0 + \omega_{s\qq})}\right]$ and $P_{\pp'}^{-} \equiv \Pr\left[\dfrac{1}{\omega - (H_{\pp'}^0 - \omega_{s\qq})}\right]$

For the first term we get
\begin{align}
    \int \dfrac{\textd\omega}{-2\pi i} \commutator{\real \Sigma^{\texteq}}{\delta\GK} = \dfrac{1}{4\pi }\int \textd\omega \commutator{\real \Sigma^{\texteq}}{\anticommutator{A^{0, \texteq}_{\pp}}{\delta h_\pp}} = -\dfrac{1}{4 }\left( I_{(1)} - \text{h.c.}\right), 
\end{align}
where
\begin{align}
    I_{(1)} &= \int \dfrac{\textd\omega}{2\pi} \anticommutator{A^{0, \texteq}_{\pp}}{\delta h_\pp}\real \Sigma^{\texteq} \nonumber\\
    & =\sum_{s\pp'}\left(\left[ (\delta h_\pp g_s^\dagger) \odot(\Piell_{s\pp\pp'} +
    \bar{\Pi}_{s\pp\pp'}h_{\pp'}^{\texteq}) \right] 
     + \delta h_\pp\left[  g_s^\dagger \odot(\Piell_{s\pp\pp'} + \bar{\Pi}_{s\pp\pp'}h_{\pp'}^{\texteq}) \right]\right)g_s, 
\end{align}
where $\Piell_{s\pp\pp'} = \Pi_{\pp\pp'}^{\text{emi}, s-\qq}l^{0, \texteq}_{-\qq}(\omega_{s-\qq})- \Pi_{\pp\pp'}^{\text{abs}, s\qq}l^{0, \texteq}_\qq(-\omega_{s\qq})$, $\bar{\Pi}_{s\pp\pp'} = \Pi_{\pp\pp'}^{\text{emi}, s-\qq}- \Pi_{\pp\pp'}^{\text{abs}, s\qq}$. Emission and adsorption $\Pi$ values are derived as the following integrals:
\begin{align}
    [\Pi_{\pp\pp'}^{\text{emi}/\text{abs},s\mp\qq}]_{ij} = \int \textd\omega \delta(\omega - \epsilon_{i\pp})\Pr\left[\dfrac{1}{\omega - (\epsilon_{j\pp'}\pm\omega_{s\qq})}\right] = \begin{cases} 0,  \text{ if } \epsilon_{i\pp} = \epsilon_{j\pp'} \pm\omega_{s\mp\qq}\\ 
   \dfrac{1}{\epsilon_{i\pp} - (\epsilon_{j\pp'}\pm\omega_{s\mp\qq})} 
   \end{cases} 
\end{align}

Similarly,
\begin{align}
    \int \dfrac{d\omega}{-2\pi i} \commutator{\real \delta \Sigma}{G^{\textK, \texteq}} = -\dfrac{1}{4 }\left( I_{(2)} - \text{h.c.}\right), 
\end{align}
where
\begin{align}
    I_{(2)} &= \sum_{s\pp'}\int \dfrac{\textd\omega}{2\pi} A_{\pp}^{0, \texteq}h_{\pp}^{\texteq}g_s^\dagger\anticommutator{P_{\pp'}^{+} - P_{\pp'}^{-}}{\delta h_{\pp'}}g_s \nonumber\\
    & = \sum_{s\pp'} \left( \left[(h_{\pp}^{\texteq}g_s^\dagger) \odot \bar{\Pi}_{s\pp\pp'} \right]\delta h_{\pp'} + \left[(h_{\pp}^{\texteq}g_s^\dagger\delta h_{\pp'})\odot\bar{\Pi}_{s\pp\pp'}\right]\right)g_s
\end{align}

Using the expression for $\SigmaK$ given in Appendix Sec. \ref{appsec:collision_eph}, we get the following for the last term:
\begin{align}
    \int \dfrac{\textd\omega}{-2\pi i} \commutator{\delta \SigmaK}{\real G^{\texteq}} = -\dfrac{1}{4}\left( I_{(3)} - \text{h.c.}\right),
\end{align}
where
\begin{align}
    I_{(3)} &= \sum_{s\pp'}\int \textd\omega  \real G^{\texteq}g^\dagger_s\left(\anticommutator{A_{\pp'}^+}{\delta h_{\pp'}l^+_\qq} - \anticommutator{A_{\pp'}^-}{\delta h_{\pp'}l^-_\qq}\right)g_s = -\left[ (g_s^\dagger \delta h_{\pp'})\odot \Piell_{s\pp\pp'}\right]g_s
\end{align}
 
\subsubsection{Diagonal theory limit} 

Let's find matrix elements of each term: 
\begin{align*}
    \commutator{\real \Sigma^{\texteq}}{\delta\GK}_{nn'}  \rightarrow-\dfrac{1}{4}\sum_{s\pp'ij}&\Bigl(\delta h_{ni\pp}g_{sij\pp\pp'}^\dagger g_{sjn'\pp\pp'}\left(\Piell_{s\pp\pp'nj} + \Piell_{s\pp\pp'ij} + h_{\pp'j}^{\texteq}(\bar{\Pi}_{s\pp\pp'nj} + \bar{\Pi}_{s\pp\pp'ij})\right) \\
    &-\delta h_{in'\pp}g_{snj\pp\pp'}^\dagger g_{sji\pp\pp'}\left(\Piell_{s\pp\pp'n'j} + \Piell_{s\pp\pp'ij} + h_{\pp'i}^{\texteq}(\bar{\Pi}_{s\pp\pp'n'j} + \bar{\Pi}_{s\pp\pp'ij})\right)
    \Bigl) \\
    \commutator{\real \delta \Sigma}{G^{\textK, \texteq}}_{nn'} \rightarrow-\dfrac{1}{4}\sum_{s\pp'ij}& g_{sni\pp\pp'}^\dagger g_{sjn'\pp\pp'}\delta h_{ij\pp'}\Bigl(h_{\pp n}^{\texteq}(\bar{\Pi}_{s\pp\pp'ni}+\bar{\Pi}_{s\pp\pp'nj})-h_{\pp n'}^{\texteq}(\bar{\Pi}_{s\pp\pp'n'i}+\bar{\Pi}_{s\pp\pp'n'j}) \Bigl) \\
    \commutator{\delta \SigmaK}{\real G^{\texteq}}_{nn'} \rightarrow +\dfrac{1}{2}\sum_{s\pp'ij}&g_{sni\pp\pp'}^\dagger g_{sjn'\pp\pp'}\delta h_{ij\pp'} \left( \bar{\Pi}_{s\pp\pp'nj} -\bar{\Pi}_{s\pp\pp'n'i} \right)
\end{align*}

In the case of diagonal $\delta h_{\pp}$: 
\begin{align*}
    \commutator{\real \Sigma^{\texteq}}{\delta\GK}_{nn'}  &\rightarrow-\dfrac{1}{2}g_{snj\pp\pp'}^\dagger g_{sjn'\pp\pp'}\sum_{s\pp'j}\Bigl(\delta h_{n\pp}\left(\Piell_{s\pp\pp'nj} + h_{\pp'j}^{\texteq}\bar{\Pi}_{s\pp\pp'nj}\right) -\delta h_{n'\pp}\left(\Piell_{s\pp\pp'n'j} + h_{\pp'j}^{\texteq} \bar{\Pi}_{s\pp\pp'n'j}\right)
    \Bigl) \\
    \commutator{\real \delta \Sigma}{G^{\textK, \texteq}}_{nn'} &\rightarrow-\dfrac{1}{2}\sum_{s\pp'j} g_{sni\pp\pp'}^\dagger g_{sjn'\pp\pp'}\delta h_{j\pp'}\Bigl(h_{\pp n}^{\texteq}\bar{\Pi}_{s\pp\pp'nj}-h_{\pp n'}^{\texteq}\bar{\Pi}_{s\pp\pp'n'j} \Bigl) \\
    \commutator{\delta \SigmaK}{\real G^{\texteq}}_{nn'} &\rightarrow +\dfrac{1}{2}\sum_{s\pp'j}g_{snj\pp\pp'}^\dagger g_{sjn'\pp\pp'}\delta h_{j\pp'} \left( \bar{\Pi}_{s\pp\pp'nj} -\bar{\Pi}_{s\pp\pp'n'j} \right)
\end{align*}

\subsection{Detailed derivation of impurity kinetic corrections}\label{appsec:kineticcorimp}
In this case, $\Sigma[G]$ is given by Eq. \ref{eq:sigmaimp}. Taking into account that $V_{\pp' \pp}^T = V_{\pp\pp'}^\dagger$, it can be seen that $\Sigma^{\textK, \dagger} = - \SigmaK$ and $\real \Sigma^{\dagger} = \real \Sigma$. As a result,
\begin{align}
    \int \dfrac{\textd\omega}{-2\pi i} \commutator{\real \Sigma^\texteq (\omega)}{\delta G^{\textK}(\omega)} = -\dfrac{1}{2}\left( I_{(1)} - h.c.\right),
\end{align}
where 
\begin{align}
    I_{(1)} &= \sum_{\pp'} \int \dfrac{\textd\omega}{2\pi} \anticommutator{\delta h_{\pp}}{A^\texteq_{\pp}} V_{\pp \pp'}  \real G_{\pp'}^{\texteq}   V_{\pp\pp'}^\dagger \nonumber \\
    & = \sum_{\pp'} \left( \left[(\delta h_{\pp}V_{\pp\pp'}) \odot  \Pi_{\pp\pp'}\right] + \delta h_{\pp}\left[V_{\pp\pp'} \odot  \Pi_{\pp\pp'}\right] \right) V_{\pp\pp'}^\dagger
\end{align}
where $[\Pi_{\pp\pp'}]_{ij}$ is
\begin{align*}
    [\Pi_{\pp\pp'}]_{ij} = \int \textd\omega \delta(\omega - \epsilon_{i\pp}) \Pr\left[\frac{1}{\omega - \epsilon_{j\pp' }} \right] = \mint{-} \frac{\delta(\omega - \epsilon_{i\pp})}{\omega - \epsilon_{j\pp'}}d\omega = \begin{cases} 0,  \text{ if } \epsilon_{j\pp'} = \epsilon_{i\pp } \\ 
   \dfrac{1}{\epsilon_{i\pp} - \epsilon_{j\pp'} } 
   \end{cases} 
\end{align*}

Similarly, we can show that: 
\begin{align}
    \int \dfrac{\textd\omega}{-2\pi i} \commutator{\delta \SigmaK (\omega)}{\real G^{\texteq}(\omega)} = -\dfrac{1}{2}\left( I_{(2)} - h.c. \right), 
\end{align}
where
\begin{align}
    I_{(2)} & = \sum_{\pp'} \int \dfrac{\textd\omega}{2\pi}  \real G_\pp^{\texteq} V_{\pp \pp'}\anticommutator{\delta h_{\pp'}}{A^\texteq_{\pp'}}V_{\pp\pp' }^\dagger \nonumber \\
    & = -\sum_{\pp'} \left( \left[ (V_{\pp \pp'}\delta h_{\pp'}) \odot \Pi_{\pp\pp'} \right] + \left[ V_{\pp \pp'} \odot \Pi_{\pp\pp'} \right]\delta h_{\pp'}\right)V_{\pp\pp' }^\dagger
\end{align}

\subsubsection{Diagonal theory limit}
We proceed similarly as before and compute matrix elements first:
\begin{align*}
     &\commutator{\real \Sigma^\texteq}{\delta G^{\textK}}_{nn'} \rightarrow -\dfrac{1}{2}\sum_{\pp'ij}   \left( \delta h_{ni\pp}V_{ij\pp\pp'}V_{jn'\pp\pp'}^\dagger( \Pi_{nj\pp\pp'} + \Pi_{ij\pp\pp'} ) -  \delta h_{in'\pp}V_{nj\pp\pp'}V_{ji\pp\pp'}^\dagger( \Pi_{n'j\pp\pp'} + \Pi_{ij\pp\pp'} )\right), \\
    &\commutator{\delta \SigmaK }{\real G^{\texteq}}_{nn'} \rightarrow  \dfrac{1}{2}\sum_{\pp'ij}   \left( V_{ni\pp\pp'} V_{jn'\pp\pp'}^\dagger \delta h_{ij\pp'} (\Pi_{nj\pp\pp'} + \Pi_{ni\pp\pp'} - \Pi_{n'j\pp\pp'} - \Pi_{n'i\pp\pp'}) \right)
\end{align*}

Then in the case of diagonal $\delta h_{\pp}$, we get
\begin{align*}
     &\commutator{\real \Sigma^\texteq}{\delta G^{\textK}}_{nn'} \rightarrow -\sum_{\pp'j} V_{nj\pp\pp'}V_{jn'\pp\pp'}^\dagger  \left( \delta h_{n\pp}\Pi_{nj\pp\pp'}  -  \delta h_{n'\pp}\Pi_{n'j\pp\pp'} \right), \\
    &\commutator{\delta \SigmaK }{\real G^{\texteq}}_{nn'} \rightarrow  \sum_{\pp'j}   \left( V_{ni\pp\pp'} V_{jn'\pp\pp'}^\dagger \delta h_{j\pp'} (\Pi_{nj\pp\pp'}  - \Pi_{n'j\pp\pp'}) \right)
\end{align*}

\subsection{Standard theory limit}\label{appsec:standardtheory}
Here we show that in the appropriate limits, our expressions corresponds to the standard, diagonal electron BTE. Below, we work on the kinetic (LHS) and collision (RHS) terms separately. Every term is multiplied by a factor of $i/2$ in order to give the results in the standard form.

\subsubsection{Kinetic side}
Under standard approximations, the kinetic corrections are neglected. The diagonal matrix elements of the commutator terms are zero. The diagonal matrix elements of the only surviving term is
\begin{equation}
    \dfrac{i}{2}\text{LHS} = -e\mathbf{E}\cdot\partial_{\pp}f^{\texteq}_{n\pp},
\end{equation}
where $e$ is the magnitude of the electronic charge.

\subsubsection{Collision side}
Here we will consider the e-ph and e-imp interactions separately. Recall that in the standard theory, the off-diagonal elements of the Wigner function are ignored.

\paragraph*{e-ph interaction} We first consider both terms $\anticommutator{\SigmaK}{A}$ and $\anticommutator{\Gamma}{\GK}$ in the case of diagonal $h_\pp$: $[h_\pp]_{nn'} = h_{n\pp}\delta_{nn'}$. Using the properties of the Hadamard product given in Eq. \ref{eq:derivinvop}, we can show that
\begin{align*}
    \dfrac{i}{2}\anticommutator{\SigmaK}{A^{0, \texteq}} &\rightarrow 
    \dfrac{i\pi}{2} \sum_{s\pp'}\left( \left[ g_s^\dagger \odot (\bar{\Delta}^{s}_{\pp\pp'} +  \Deltaell^{s}_{\pp\pp'} h_{\pp'})\right]g_s + h.c. \right),
\end{align*}

\begin{align*}
\dfrac{i}{2}{\anticommutator{\Gamma}\GK}&\rightarrow 
\dfrac{i\pi}{2}\sum_{s\pp'}\left( \left[ g^\dagger_s\odot (h_\pp(\bar{\Delta}^{s}_{\pp\pp'}h_{\pp'}+ \Deltaell^{s}_{\pp\pp'})) \right]g_s + h.c.\right)
\end{align*} 

Finally, the e-ph collision integral becomes
\begin{align*}
\dfrac{i}{2}\anticommutator{\SigmaK}{A^{0, \texteq}} - \dfrac{i}{2}{\anticommutator{\Gamma}\GK}&\rightarrow \dfrac{i\pi}{2}\sum_{s\pp'}\left(\left[g^\dagger_s\odot (\bar{\Delta}^{s}_{\pp\pp'} +  \Deltaell^{s}_{\pp\pp'}h_{\pp'} - h_\pp(\bar{\Delta}^{s}_{\pp\pp'}h_{\pp'} + \Deltaell^{s}_{\pp\pp'})) \right]g_s + h.c.\right).
\end{align*} 

Next, we consider the matrix elements of the above expression. After some transformations and using the fact, that $l^{0, \texteq}_{s\qq}(-\omega) = - l^{0, \texteq}_{s\qq}(\omega)$, we get
\begin{align*}
(nn')&: \dfrac{i\pi}{2}\sum_{s\pp'}\sum_ig^\dagger_{sni}g_{sin'}\biggl( Y_{ni} + Y_{n'i}\biggl),  
\end{align*} 
where 
$Y_{ni} = [\Delta^{\text{emi}, s-\qq}_{\pp\pp'}]_{ni}\left[1 + l^{0}_{s-\qq}h_{i\pp'} - l^{0}_{s-\qq}h_{n\pp} - h_{i\pp'}h_{n\pp}\right]  + [\Delta^{\text{abs}, s\qq}_{\pp\pp'}]_{ni}\left[-1 + l^{0}_{s\qq}h_{i\pp'} - l^{0}_{s\qq}h_{n\pp} + h_{i\pp'}h_{n\pp}\right]$

In equilibrium $Y_{ni} = 0$ due to the identities given in Eq. \ref{eq:wigneridentities}. Using this, we finally get
\begin{align}
    \dfrac{i}{2}\text{RHS}_{\text{e-ph}} &= 2\pi \sum_{sn'\pp'}|g^{snn'}_{\pp\pp'}|^{2} \nonumber \\
    &\qquad\qquad\Big\{\left[(1 - f_{n\pp})f_{n'\pp'}(1 + n^{0}_{s\qq}) - f_{n\pp}(1 - f_{n'\pp'})n^{0}_{s\qq}\right]\delta(\epsilon_{n\pp} - \epsilon_{n'\pp'} + \omega_{s\qq}) \nonumber \\
    &\qquad\qquad+ \left[(1 - f_{n\pp})f_{n'\pp'}n^{0}_{s-\qq} - f_{n\pp}(1 - f_{n'\pp'})(1 + n^{0}_{s-\qq})\right]\delta(\epsilon_{n\pp} - \epsilon_{n'\pp'} - \omega_{s-\qq}) \Big\} \nonumber \\
    &\overset{\text{linearization}}{=} 2\pi \sum_{sn'\pp'}|g^{snn'}_{\pp\pp'}|^{2} \nonumber \\
    &\qquad\qquad\Big\{\left[(1+ n^0_{s\qq} - f^0_{n\pp})\delta f_{n'\pp'} - (n^0_{s\qq} + f^0_{n'\pp'})\delta f_{n\pp}\right]\delta(\epsilon_{n\pp} - \epsilon_{n'\pp'} + \omega_{s\qq}) \nonumber \\
    &\qquad\qquad+ \left[(n^0_{s-\qq} + f^0_{n\pp})\delta f_{n'\pp'} - (1  + n^{0}_{s-\qq} - f^{0}_{n'\pp'})\delta f_{n\pp} \right]\delta(\epsilon_{n\pp} - \epsilon_{n'\pp'} - \omega_{s-\qq}) \Big\}
\end{align}

Where in the last line we substituted $f_{n\pp} = f_{n\pp}^0 + \delta f_{n\pp}$.





\paragraph*{e-imp interaction} Collecting the diagonal parts of the linearized expressions for the in- and out-scattering terms from Sec. \ref{sec:matrixBTE}, we straightforwardly get
\begin{equation}
    \dfrac{i}{2}\text{RHS}_{\text{e-imp}} = 2\pi\sum_{n'\pp'}\left|V_{nn'\pp\pp'}\right|^{2}\left( \delta f_{n'\pp'} - \delta f_{n\pp} \right)\delta(\epsilon_{n\pp} - \epsilon_{n'\pp'}).
\end{equation}

\section*{Acknowledgments} This work was funded by the Deutsche Forschungsgemeinschaft (DFG, German Research
Foundation) through the Emmy Noether research grant
(Grant No. 534386252). N. H. P. acknowledges fruitful discussions with Gianluca Stefanucci and Enrico Perfetto.

\section*{Data availability} All data supporting this work is already included in the manuscript.

\bibliography{refs}

\end{document}